\documentclass[12pt]{article}
\begin{document}

\title{Fragmentation of multiply charged simple metal clusters in
liquid-drop stabilized jellium model}
\author{M. Payami}
\maketitle
\begin{center}
{\it Center for Theoretical Physics and Mathematics, Atomic
Energy Organization of Iran,\\ P.~O.~Box 11365-8486, Tehran, Iran}
\end{center}

\begin{abstract}

In this work, we have used the liquid-drop model in the context
of stabilized jellium model, to study the stability of $Z$-ply
charged metal clusters of different species against fragmentation.
We have shown that on the one hand, singly ionized clusters are
stable against any spontaneous fragmentation, and on the other
hand, the most favored decay process for them is atomic
evaporation. However, multiply charged clusters of sufficiently
small sizes may undergo spontaneous decay via fission processes.
Comparing the results for different species show that for fixed
$N$, the lower electron density metal clusters can accommodate
more excess charges before their Coulomb explosions. This
comparison also shows that, for fixed $Z$, the atomic evaporation
which is the most favored decay mechanism for sufficiently large
clusters, takes place at lower $N$s for lower electron density
clusters.
\end{abstract}
\newpage

\section{Introduction}
\label{sec1}

Study of the stability of charged metal clusters against
fragmentation is important in nano-sized systems. In this work,
we have studied the binary fragmentation of $N$-atom $Z$-ply
charged ${\cal M}_N^Z$ clusters in the framework of the
liquid-drop \cite{Brack93} model (LDM) with stabilized
\cite{pertran} jellium model (SJM) coefficients for the energies.
Here, $Z$=1, 2, 3, 4; and ${\cal M}$ covers the species Al, Ga,
Li, Na, K, and Cs. Here, in the case of Al, when $N$ is a
multiple of 3, it corresponds to real Al$_\nu$ cluster with
$\nu=N/3$. The clusters are assumed to be spherical with sizes
determined by $R=N^{1/3}r_s^B$ with $N\le 100$. The quantity
$r_s^B$ is the Wigner-Seitz radius of the electrons in the bulk
metal. The possible decay channels of ${\cal M}_N^Z$ are

\begin{equation}\label{eq1}
  {\cal M}_N^Z\to {\cal M}_{N-p}^{Z-Z_1}+{\cal M}_p^{Z_1}
\end{equation}
Here, $p$ is positive integer as $N$, and  $0\le Z_1\le [Z/2]$.
Different processes are defined by different values of $Z_1$. The
process in which $Z_1=0$ (one of the fragments is neutral), is
called evaporation. The processes in which $0<Z_1=Z/2$ are called
charge-symmetric fission, and the charge-asymmetric fission is
defined by $0<Z_1<[Z/2]$. For any given set of values {$N$, $Z$,
$Z_1$}, different values of $p$ define different decay channels.
In a particular decay channel of the evaporation processes, the
negativity of the difference between total energies before and
after the fragmentation,

\begin{equation}\label{eq2}
D^Z(N,p)=E^Z(N-p) + E^0(p) - E^Z(N),
\end{equation}
is sufficient for the decaying in that particular channel. In the
above equation, $E^Z(N)$ and $E^0(N)$ are the total energies of
$Z$-ply ionized and neutral $N$-atom clusters, respectively.
However, in fission processes a negative value for the difference
energy is not a sufficient condition for the decay of the parent
cluster. This is because, the competition between the short-range
surface tension and the long-range repulsive Coulomb force may
give rise to a fission barrier. The heights of the fission
barriers are calculated using the two-spheres
approximation\cite{Naher}. In Fig. \ref{fig1}, the fission of a
$Z$-ply charged $N$-atom cluster into two clusters of respective
sizes $N_1$, $N_2=N-N_1$ and respective charges $Z_1$,
$Z_2=Z-Z_1$ is schematically shown. $Q_f$ is the energy release,
$B_c$ is the fusion barrier (or Coulomb barrier) which is the
maximum energy of the Coulomb interaction of two
positively-charged conducting spheres, taking their
polarizabilities into account. $B_f$ is the fission barrier
height which is defined as

\begin{equation}\label{eq3}
B_f=-Q_f+B_c.
\end{equation}

The Coulomb interaction energy, $E_c$, as a function of their
separations, $d$, for two charged metallic spheres can be
numerically calculated using the classical method of image
charges \cite{Naher}.

The most favored decay channel in evaporation processes is
defined as the channel for which the dissociation energy assumes
its minimum value

\begin{equation}\label{eq4}
D^Z(N,p^*)=\min_p\left\{D^Z(N,p)\right\},
\end{equation}
and the most favored decay channel in fission processes is
defined as the channel for which the fission-barrier height
assumes its minimum value,

\begin{equation}\label{eq5}
B_f(N,p^*)=\min_p\left\{B_f(N,p)\right\}.
\end{equation}

The energy of an $N$-atom $Z$-ply charged cluster in the LDM is
given \cite{seidl97,payami_aw} by

\begin{equation}\label{eq6}
  E^Z(N)=E^0(N)+Z(W+\frac{c}{R})+\frac{Z^2e^2}{2(R+a)},
\end{equation}
in which $W$, $c$, $R$, and $e$ are the work function of the bulk
metal, the finite-size correction to the work function, the
radius of the cluster, and the electron charge, respectively. For
simplicity, the position of the centroid of excess charge, $a$,
is neglected in our calculations. $E^0(N)$ is the energy of a
neutral $N$-atom cluster in the LDM, which is given by

\begin{equation}\label{eq7}
  E^0(N)=\varepsilon N+4\pi (r_s^B)^2\sigma N^{2/3} + 2\pi r_s^B
  \gamma N^{1/3}.
\end{equation}
In Eq. (\ref{eq7}), the quantities $\varepsilon$, $\sigma$, and
$\gamma$ are total energy per electron of the bulk, surface
energy, and curvature energy, respectively. We have calculated
the quantities $c$, $W$, $\sigma$, and $\gamma$ by fitting to the
self-consistent Kohn-Sham \cite{kohnsham} results in the SJM for
different $r_s$ values \cite{payami_aw}. However, $\varepsilon$ is
calculated using the SJM energy expression for the bulk system
(Eq. (1) of Ref. [\ref{payamiJPC01}]).

\section{Results and discussion}
\label{sec2}

Using the method of image charges, we have calculated the Coulomb
interaction energy of two charged metallic spheres, taking their
polarizabilities into account. The calculations show that the
maximum of the interaction energy, $B_c$, is achieved for a
separation $d_0\ge R_1+R_2$. Fig. \ref{fig2}(a) shows the Coulomb
interaction energy of an $N_1$-atom cluster with another
$N_2$-atom of respective excess charges $Z_1=3$ and $Z_2$=1. The
radii of the clusters are calculated from $R=N^{1/3}r_s^B$, as in
the SJM. In this figure, we have taken $r_s^B$=2.07 which
correspond to bulk Al. When both $N_1$ and $N_2$ are multiples of
3, the results correspond to real Al clusters. The value at the
maximum specifies the quantity $B_c$, and the position of the
maximum, $d_0$, is the separation between the centers of the two
spheres. As is seen, for fixed values of charges, $B_c$ is the
highest when the sizes are equal. In Fig. \ref{fig2}(b), the
situation is shown for equal charges $Z_1=Z_2=2$. In this case,
when the sizes are equal, $B_c$ is maximum as before but here,
$d_0=R_1+R_2$. An other feature shown in Fig. \ref{fig2}(c) is
that, when both charges and sizes are asymmetric, $B_c$ is higher
if the smaller charge corresponds to the smaller cluster. In Fig.
\ref{fig2}(d), we have compared the Coulomb interaction energies
of two Al metallic spheres with respective sizes $N_1=2$ and
$N_2=18$ for different charges. It is seen that keeping the sizes
fixed but increasing the charge on any one of them, without
changing the charge on the other one, increases the height of the
maximum.

In our calculations we need the values of the Coulomb interaction
energy at the maxima points, i.e., the $B_c$s. In Fig.
\ref{fig3}(a), we have compared the $B_c$ values for equally
charged $Z_1=Z_2=1$, Al metallic particles but with different
radii $R_1=2.07 N_1^{1/3}$ and $R_2=2.07 N_2^{1/3}$. This figure
shows that when both of the sizes are small, the barrier is
higher than the case when at least one of them is larger. In Fig.
\ref{fig3}(b), we have compared the $B_c$s for singly ionized
pairs Al, Na, and Cs with different values of $N_1$ and $N_2$. In
the aluminum case, since the radii of the spheres are smaller
than those of Na and so on, their polarizabilities are thereby
smaller and thus the Al values lie above that of Na, and so on.
Now, if the charge on one of the clusters is increased without
changing that of the other one, the arrangement of the curves
would not change but they shift upward according to Fig.
\ref{fig2}(d).

In the energy calculations of Eq. (\ref{eq6}) we use the results
obtained for $c$ and $W$ in Ref. [\ref{payami_aw}]. The values of
$\varepsilon$ are calculated using Eq. (1) of Ref.
[\ref{payamiJPC01}] for unpolarized case. $\sigma$ and $\gamma$
in Eq. (\ref{eq7}) are obtained by fitting the results of the
self-consistent solutions of the Kohn-Sham equations in the SJM
for neutral clusters of sizes $N\le 100$, and different $r_s$
values. The values of $r_s^B$ for Al, Ga, Li, Na, K, and Cs are
2.07, 2.19, 3.28, 3.99, 4.96, and 5.63, respectively. In Figs.
\ref{fig4}(a)-(c) we have plotted the SJM values of $\varepsilon$,
$\sigma$, and $\gamma$, as functions of $r_s$, respectively.

To show the variations of the dissociation energy, $D^Z(N,p)$,
for different charging and different evaporation channels, we
have plotted, in Figs. \ref{fig5}(a)-(b), the quantities
$D^{1+}(N,p)$ and $D^{4+}(N,p)$, respectively, as functions of the
neutral fragment size, $p$, for Li. As is seen in Fig.
\ref{fig5}(a), for small $p$ values $D^{1+}(N,p)$ has an
increasing behavior, then after passing a maximum, it changes to a
decreasing behavior, and finally, for large $p$ values it
increases again. This behavior has its roots in the trade off
between the surface energy (the volume energy term does not
change in the fragmentation process) and the last term of Eq.
(\ref{eq6}). For small $p$, the surface area difference

\begin{equation}\label{eq8}
 \Delta S(R_1,R)=4\pi\left[R_1^2+(R^3-R_1^3)^{2/3}-R^2\right],
\end{equation}
increases, and then for intermediate $p$ values, it assumes a
maximum, and finally decreases. The increasing behavior at the
tail is because, with increasing $p$, the size of the singly
ionized fragment decreases and hence, beyond a certain $p$ value,
the last term in Eq. (\ref{eq6}) dominates the surface energy
term. However, beyond a $Z$ value for the parent cluster, the
last term of the Eq. (\ref{eq6}) dominates the surface energy
term for smaller $p$ values before $D^Z(N,p)$ reaches its maximum
as in the case of Fig. \ref{fig5}(b). In fact, we have found that
for Li$_{70}^Z$ this transition takes place at a point
$1.6<Z_t<1.7$. The value of $Z_t$ depends on $N$. This dependence
is clearly shown in Fig. \ref{fig5}(a) by the fact that
$D^{1+}(20,p)$ and $D^{1+}(10,p)$ are increasing functions of p
in the whole range.

The dissociation energy $D^Z(N,p)$ for small $p/N$ takes the form

\begin{equation}\label{eq9}
  D^Z(N,p)\to 2\pi\gamma r_s p^{1/3}+4\pi\sigma r_s^2 p^{2/3},
\end{equation}
which specifies the most favored channel in an evaporation
process with $p^*=1$. This means that, in evaporation processes,
the parent charged cluster prefers to evaporate single atoms than
larger neutral clusters. This fact is also numerically shown in
Figs. \ref{fig5}(a)-(b). On the other hand, the positivity of
$D^Z(N,p)$ in the whole range implies that the parent charged
cluster is stable against any spontaneous evaporation.

In Fig. \ref{fig6}, we have plotted the most favored dissociation
energies, $D^Z(N,p^*)$, of Na clusters as functions of the parent
size, $N$, for different $Z$ values. The results show that for
large enough clusters, the dissociation energy is independent of
$Z$, which is consistent with Eq. (\ref{eq9}). However, for small
clusters, the dissociation energy increases with charge.

Figures \ref{fig7}(a)-(b) compares the most favored dissociation
energies of different species for $Z=1$ and $Z=4$, respectively.
It is seen that detachment of a single atom from a lower electron
density cluster is easier, i.e., it needs less energy. Comparing
these two figures also shows that, although the hierarchy is the
same, the dissociation energies are shifted to higher energies as
we go from $Z=1$ to $Z=4$.

Now, we consider the fission processes. We first consider the
following two cases:

\begin{equation}\label{eq10}
  {\rm Ga}_N^{4+}\to {\rm Ga}_{N-p}^{2+}+{\rm Ga}_p^{2+},
\end{equation}

\begin{equation}\label{eq11}
  {\rm Ga}_N^{4+}\to {\rm Ga}_{N-p}^{3+}+{\rm Ga}_p^{1+}.
\end{equation}

Processes (\ref{eq10}) and (\ref{eq11}) describe charge-symmetric
and charge-asymmetric fissions, respectively. In Figs.
\ref{fig8}(a) and \ref{fig8}(b) we have plotted their respective
fission barriers, $B_f^{Z,Z_1}(N,p)$, for $Z_1=2,1$ as functions
of $p$, for different sizes of the parent. As is seen in both of
these figures, there exists a minimum size for the parent,
$N_{min}^{Z,Z_1}$, beyond which the fission barrier height is
positive in all possible channels of that particular process.
That is, the parent cluster ${\cal M}^Z_N$ with $N\ge
N_{min}^{Z,Z_1}$ is stable against any spontaneous fission via
that particular process.

One of the significant differences in the charge-symmetric and
charge-asymmetric fission is that [as seen in Figs. \ref{fig8}(a)
and \ref{fig8}(b)], in the most favored channel, the products
have more or less the same sizes in the former case; whereas, in
the latter case the less charged fragment assumes quite smaller
sizes. That is, the charge-symmetric fission proceeds mostly via
mass-symmetric channel, and the charge-asymmetric fission does it
via mass-asymmetric channel. In Fig. \ref{fig8}(c), we compare
the most favored sizes, $p^*$, for the charge-symmetric and
charge-asymmetric fission of Na$^{4+}_N$. It is seen that in the
charge-symmetric case, $p^*$ increases with $N$ whereas, in the
charge-asymmetric case it remains more or less constant.

To decide whether a cluster with a given size and charge, ${\cal
M}^Z_N$, is stable against any spontaneous decay, one should
compare $N$ with the quantity

\begin{equation}\label{eq12}
  N_{min}^Z=\max_{Z_1}\left\{N_{min}^{Z,Z_1}\right\}.
\end{equation}
If $N\ge N_{min}^Z$, then the cluster ${\cal M}^Z_N$ is stable
against any spontaneous decay. To show the values of
$N_{min}^{Z,Z_1}$ schematically, we plot the most favored fission
barriers, $B_f^{Z,Z_1}(N,p^*)$ for all possible $Z_1$ values. The
evaporation processes need not be considered because, as was
shown before, in the LDM there is no spontaneous decay via
evaporation. However, in order to determine at what size which
process is dominant, we include the most favored evaporation
process as well. It should be mentioned that, in the
self-consistent Kohn-Sham results \cite{payami_rigrlx}, because
of the shell effects, the spontaneous evaporation is also
possible.

In Fig. \ref{fig9}, we have compared the most favored fission
barriers, $B_f^{4,Z_1}(N,p^*)$, of Ga$^{4+}_N$ for $N\le 150$ and
$Z_1=1,2$ with the most favored dissociation energy, $D^4(N,p^*)$.
We have labeled the intersection points of the curves with each
other and with the horizontal $N$ axis by A, B, C, D, and E. At
point A, the charge-symmetric and charge-asymmetric fission start
their competition; thus we name the corresponding size as
$N_{s-a}$. For clusters ${\cal M}^{4+}_N$ with $N<N_{s-a}$, the
charge-symmetric fission is dominant whereas for larger sizes
charge-asymmetric one dominates. We name the size corresponding
to the point B as $N_{min}^{4,2}$ because for sizes greater than
this there is no charge-symmetric spontaneous fission. However,
spontaneous fission via charge-asymmetric process still persists
for $N_{min}^{4,2}<N<N_{min}^{4,1}$. Beyond the size
$N_{min}^{4,1}$ which corresponds to the point C, the spontaneous
charge-asymmetric fission also stops. Our results show that, for
small $r_s$ values, $N_{min}^{4,2}\le N_{min}^{4,1}$ whereas, for
$r_s>4.5$, the reverse inequality is at work (as in the cases of
K and Cs).

According to Eq. (\ref{eq12}), if $N>N_{min}^Z$, then any
spontaneous fission stops and the cluster ${\cal M}^{4+}_N$ would
be stable against any spontaneous decay. At sizes $N_{eva}^{4,2}$
and $N_{eva}^{4,1}$ which correspond, respectively, to the points
D and E, competition of the evaporation with charge-symmetric and
charge-asymmetric fission starts, respectively. Beyond the size

\begin{equation}\label{eq13}
  N_{eva}^Z=\max_{Z_1}\left\{N_{eva}^{Z,Z_1}\right\},
\end{equation}
the dominant decay process is evaporation. Our results show that,
at least for $r_s\le 7$, the inequality $N_{eva}^{4,2} <
N_{eva}^{4,1}$ is satisfied. When $Z=3$, the decay processes
consist of evaporation and charge-asymmetric fission whereas, for
$Z=2$, the fission is a charge-symmetric one.

In Figs. \ref{fig10}(a) and (b), we compare the most favored
fission barrier heights $B_f^{3,1}(N,p^*)$ and $B_f^{2,1}(N,p^*)$
for different species, respectively. The figures indicate that,
for large enough clusters, the fission barriers are higher for
higher electron density metals.

In Fig. \ref{fig11}(a), we compare the smallest stable sizes,
$N_{min}^Z$ for different species. Looking at the figure, one
concludes that, for fixed $N$, the higher electron density metal
clusters (say Al) has less capacity for charging than the lower
density metal clusters (say Cs). This behavior is observed for
all $1<Z\le 4$.

To show the size beyond which evaporation dominates all decay
processes, we plot $N_{eva}^Z$ with $Z=2,3,4$ for all species in
Fig. \ref{fig11}(b). It is seen that, for fixed charge, $Z$, the
evaporation process dominates at smaller $N$ values for lower
electron density metal clusters. The trends in Figs.
\ref{fig11}(a) and \ref{fig11}(b) are seen to be the same.

Finally, in Fig. \ref{fig11}(c), we compare the $N_{s-a}$, the
size at which the competition between the charge-symmetric and
charge-asymmetric fission starts. This size decreases from Al to
Li, and then increases from Li to K, and finally decreases from K
to Cs. In the plot, we have taken into account the decimals for
the intersection points which can be rounded to the next nearest
integer.

\section{Summary and conclusion}
\label{sec3}

In this work, we have studied the stability of $Z$-ply charged
($Z=1,2,3,4$) metal clusters of species Al, Ga, Li, Na, K, and Cs
using the liquid-drop model with stabilized jellium model
energies. Fragmentation of clusters into smaller ones can proceed
via evaporation or fission. Our results show that, in the LDM,
any charged cluster is stable against spontaneous evaporation,
and the most favored channel of induced fragmentation depends on
the size and the charge of the parent cluster. Sufficiently small
multiply charged clusters, however, may undergo spontaneous
fragmentation via different fission processes. We have obtained,
for each species, the smallest size that a $Z$-ply charged
cluster is stable against any spontaneous fragmentation. We have
also shown that, for sufficiently large clusters the most favored
decay channel is atomic evaporation. Comparing the results for
different species show that, for fixed $N$, the lower electron
density clusters have higher capacity for charging. Results also
show that, for fixed amount of charge, atomic evaporation
dominates at lower $N$s for lower electron density metal clusters.

\newpage

\begin{figure}
\caption{Fission barrier in the two-spheres approximation. The
parent $N$-atom $Z$-ply charged cluster decays into two clusters
of sizes $N_1$ and $N-N_1$, with charges $Z_1$ and $Z_2$,
respectively. } \label{fig1}
\end{figure}

\begin{figure}
\caption{Coulomb interaction energy of two charged metal spheres,
in electron volls, as function of distance. The value of the
maximum is $B_c$, and the location of the maximum is at $d_0$.
(a)-the spheres have charges 3 and 1, (b)-the spheres have equal
charges $Z=2$, (c)-the charges of the two spheres are exchanged,
and (d)-the sizes are kept fixed but the charges are changed. }
\label{fig2}
\end{figure}

\begin{figure}
\caption{Coulomb barrier height, $B_c$, in electron volts, (a)-for
two equally charged but different sized pairs of the same
species, (b)-comparison of $B_c$ for different species of equally
charged pairs with one of the pair sizes kept fixed. }\label{fig3}
\end{figure}

\begin{figure}
\caption{Functions of $r_s$, (a)-total SJM energy per electron in
the bulk metal, (b)-the surface energy obtained from fitting to
the Kohn-Sham results, and (c)-the curvature energy obtained from
fitting. }\label{fig4}
\end{figure}

\begin{figure}
\caption{Dissociation energy, in electron volts, for different
evaporation channels and different parent cluster
sizes.(a)-singly ionized, (b)-4-ply ionized Li clusters.}
\label{fig5}
\end{figure}

\begin{figure}
\caption{Na dissociation energy, in electron volts, as function of
the parent cluster size, for the most favored channel of
different evaporation processes }\label{fig6}
\end{figure}

\begin{figure}
\caption{Most favored channel dissociation energies, in electron
volts, of different species for (a)-singly ionized clusters,
(b)-4-ply ionized clusters. }\label{fig7}
\end{figure}

\begin{figure}
\caption{Fission barrier heights of 4-ply ionized Ga clusters, in
electron volts, for different parent sizes and fission channels of
(a)-symmetric, (b)-asymmetric processes. }\label{fig8}
\end{figure}

\begin{figure}
\caption{Barrier energies, in electron volts, in the most favored
channel of different sizes of 4-ply ionized Ga clusters.
$Z_1=1,2$ correspond to asymmetric and symmetric fission,
respectively, while the $Z_1=0$ case correspond to evaporation
process. } \label{fig9}
\end{figure}

\begin{figure}
\caption{Most favored fission barrier heights, in electron volts,
of different species for (a)- 3-ply ionized, and (b)-doubly
ionized clusters. } \label{fig10}
\end{figure}

\begin{figure}
\caption{Functions of $r_s$, (a)-the minimum stable size for
different charges, (b)-the sizes at which evaporation dominates
fission, and (c)-the sizes at which the symmetric and asymmetric
fissions start their competitions. } \label{fig11}
\end{figure}


\newpage
\begin{thebibliography}{99}

\bibitem{Brack93}
M. Brack, Rev. Mod. Phys. {\bf 65}, 677 (1993). \label{Brack93}
\bibitem{pertran}
J. P. Perdew, H. Q. Tran, and E. D. Smith, Phys. Rev. B {\bf 42},
11627 (1990). \label {pertran}
\bibitem{Naher}
U. N\"aher, S. Bj{\o}rnholm, S. Frauendorf, F. Garcias, and C.
Guet, Phys. Rep. {\bf 285}, 245 (1997) and references therein.
\label{Naher}
\bibitem{seidl97}
M. Seidl, J. P. Perdew, M. Brajczewska, and C. Fiolhais, Phys.
Rev. B {\bf 55}, 13288 (1997). \label{seidl97}
\bibitem{payami_aw}
M. Payami, arXiv.org: cond-mat/0212160. \label{payami_aw}
\bibitem{kohnsham}
W. Kohn and L. J. Sham, Phys. Rev. {\bf 140}, A1133 (1965).
\label{kohnsham}
\bibitem{payamiJPC01}
M. Payami, J. Phys.: Condens. Matter {\bf 13}, 4129 (2001).
\label{payamiJPC01}
\bibitem{payami_rigrlx}
M. Payami, arXiv.org: cond-mat/0305600. \label{payami_rigrlx}

\end{thebibliography}
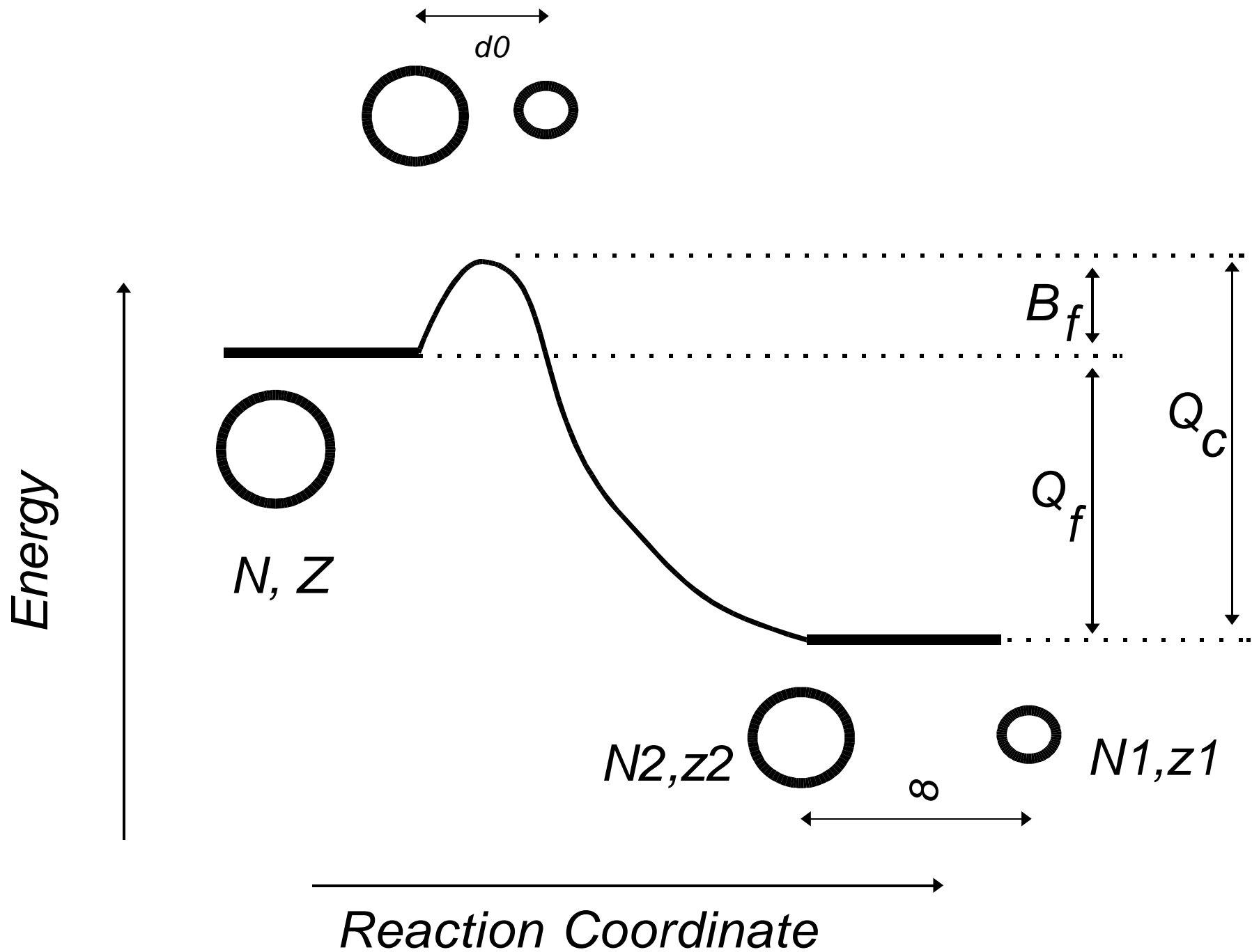
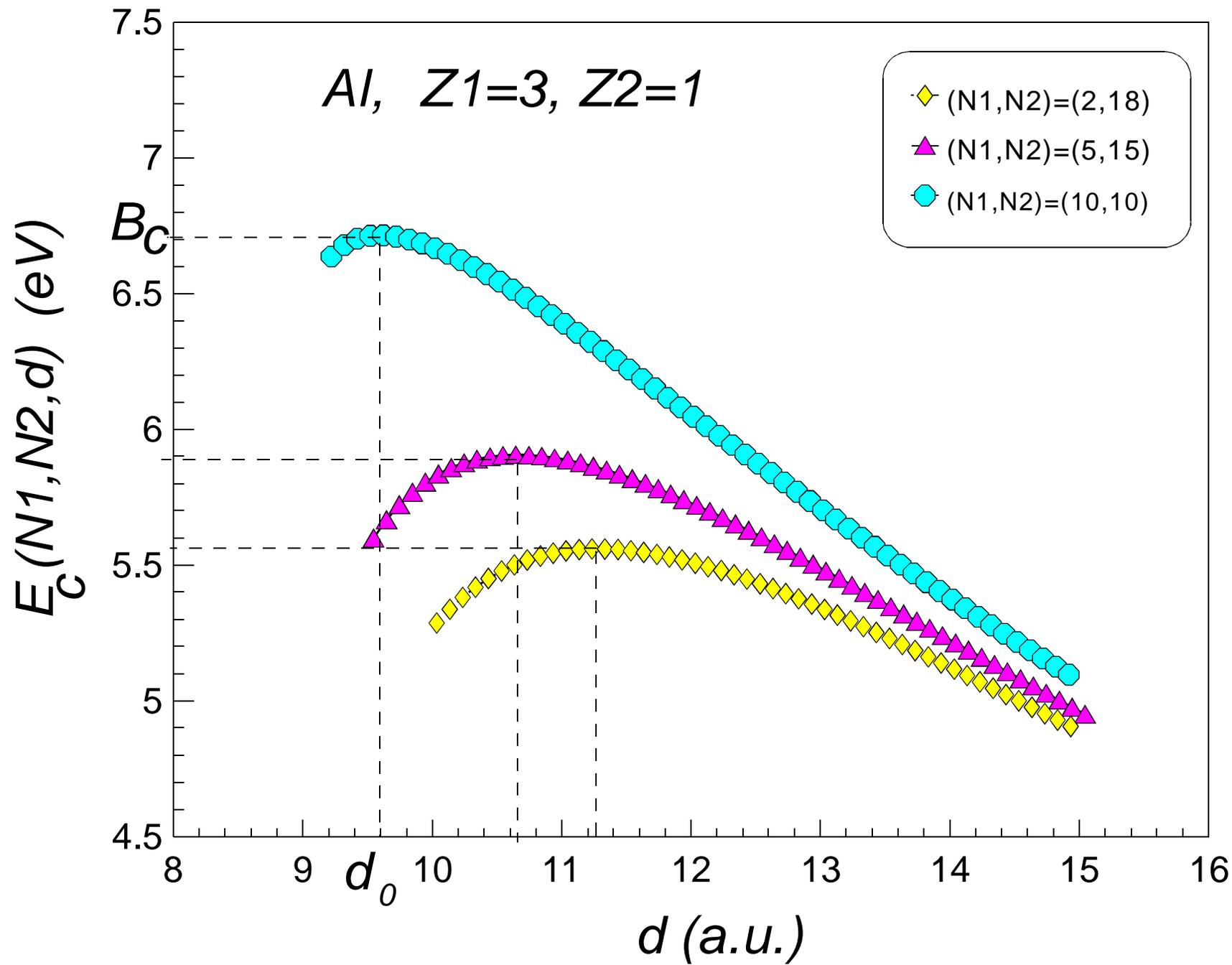
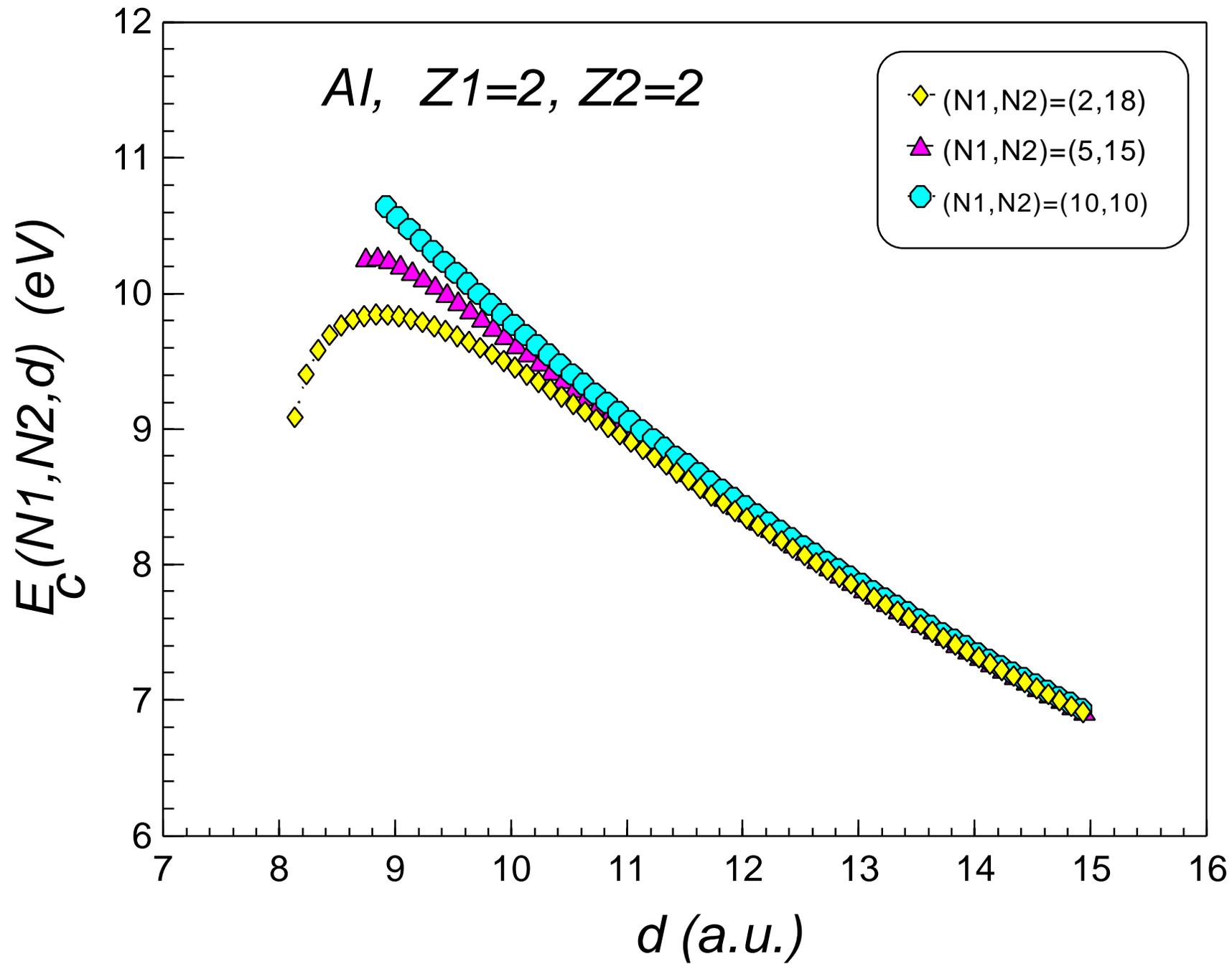
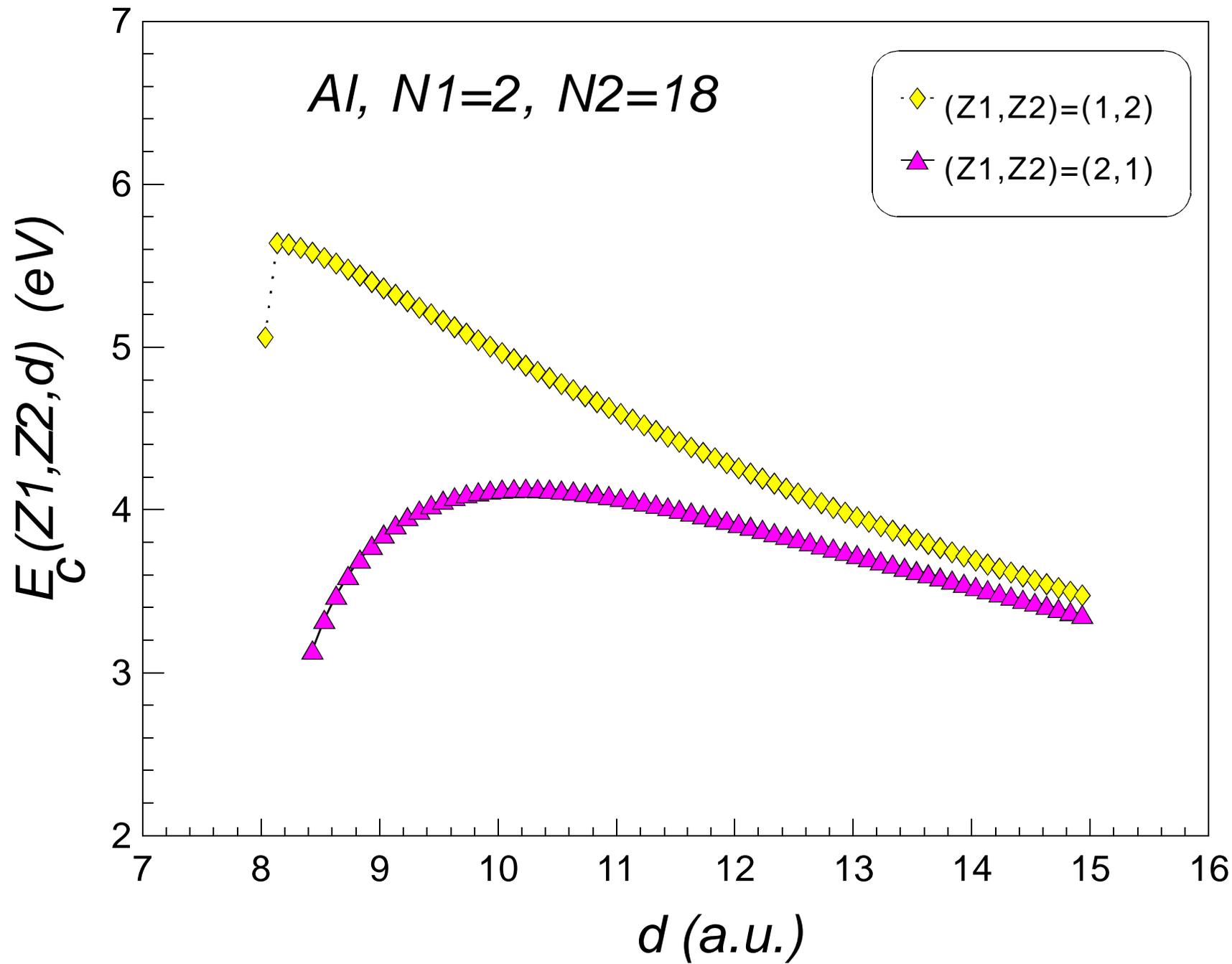
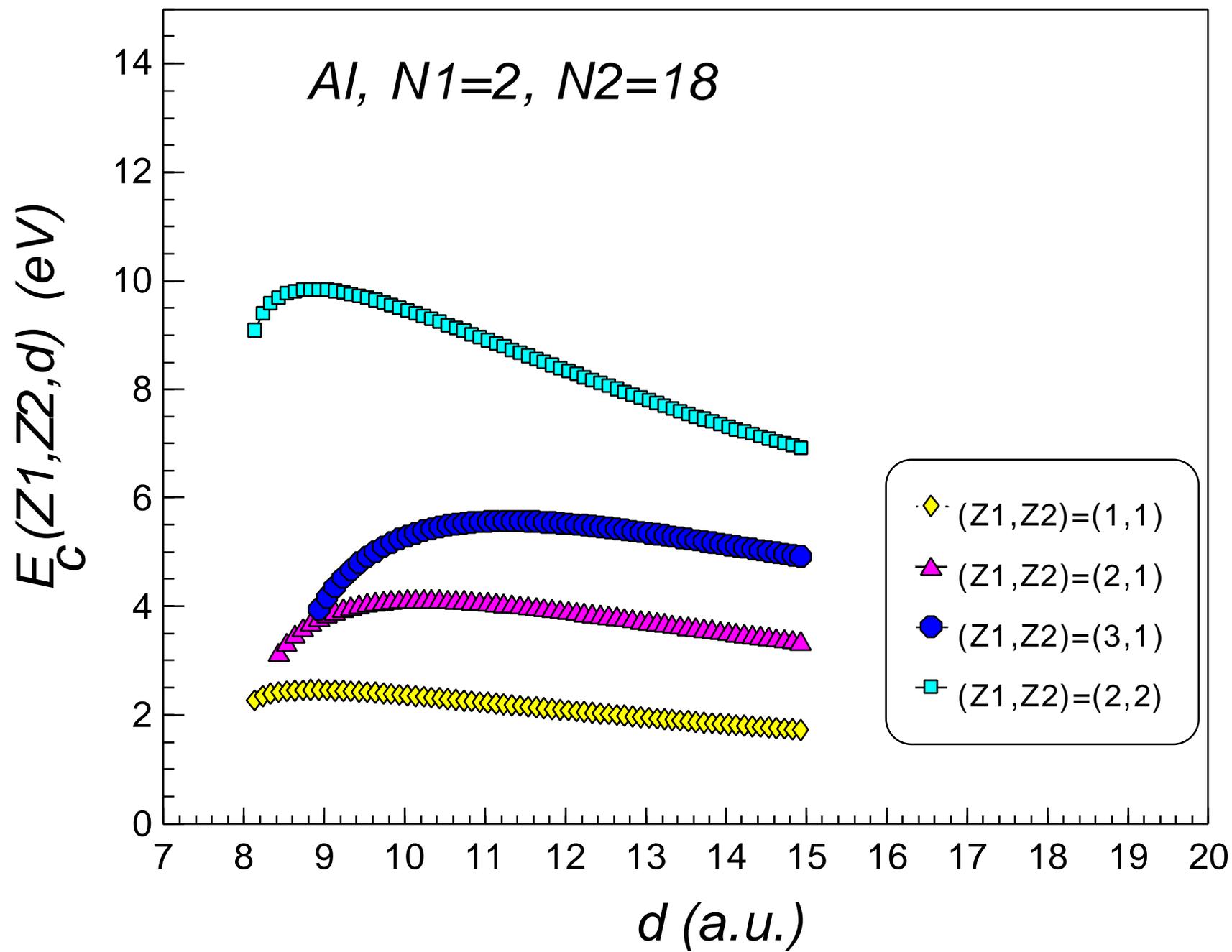
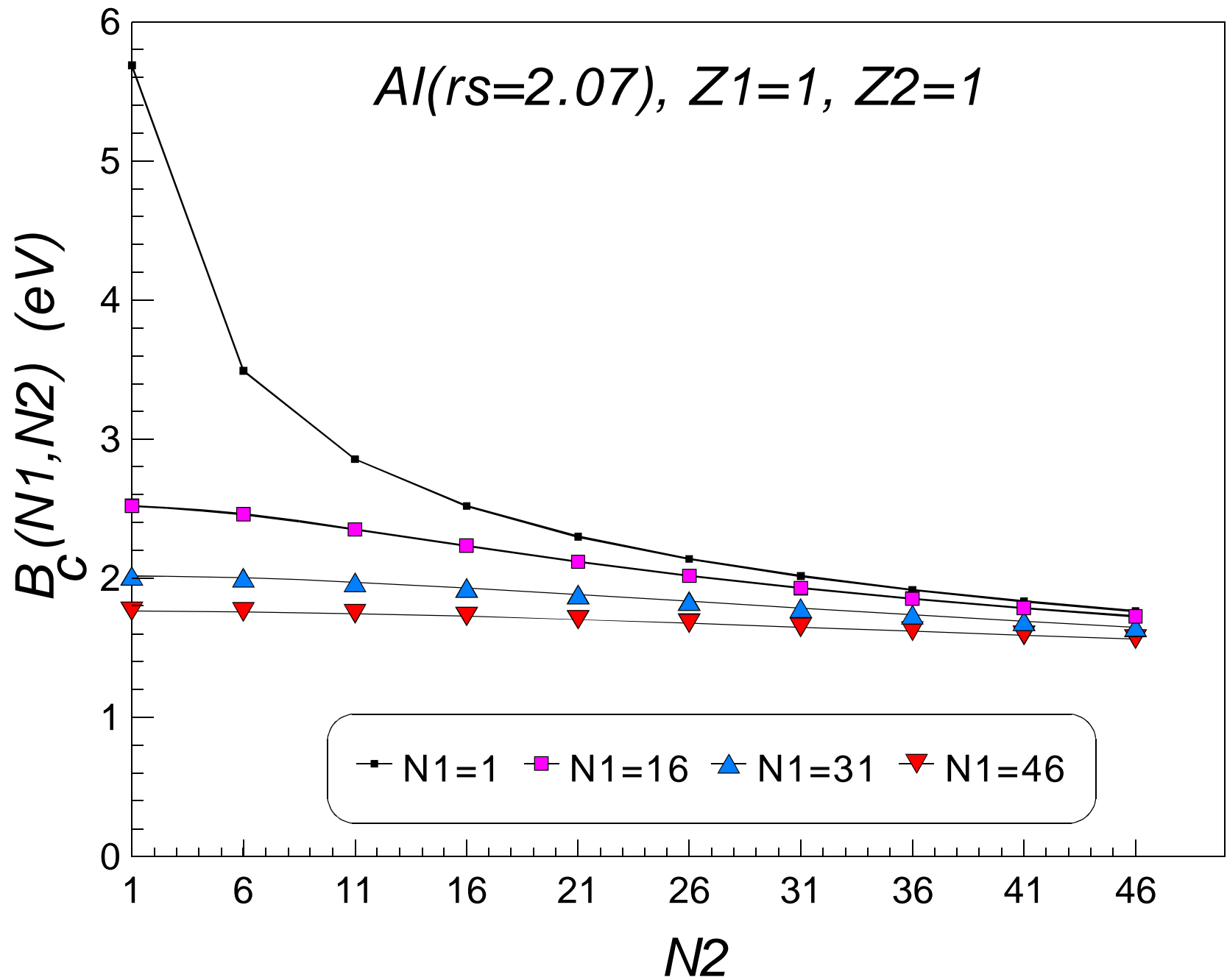
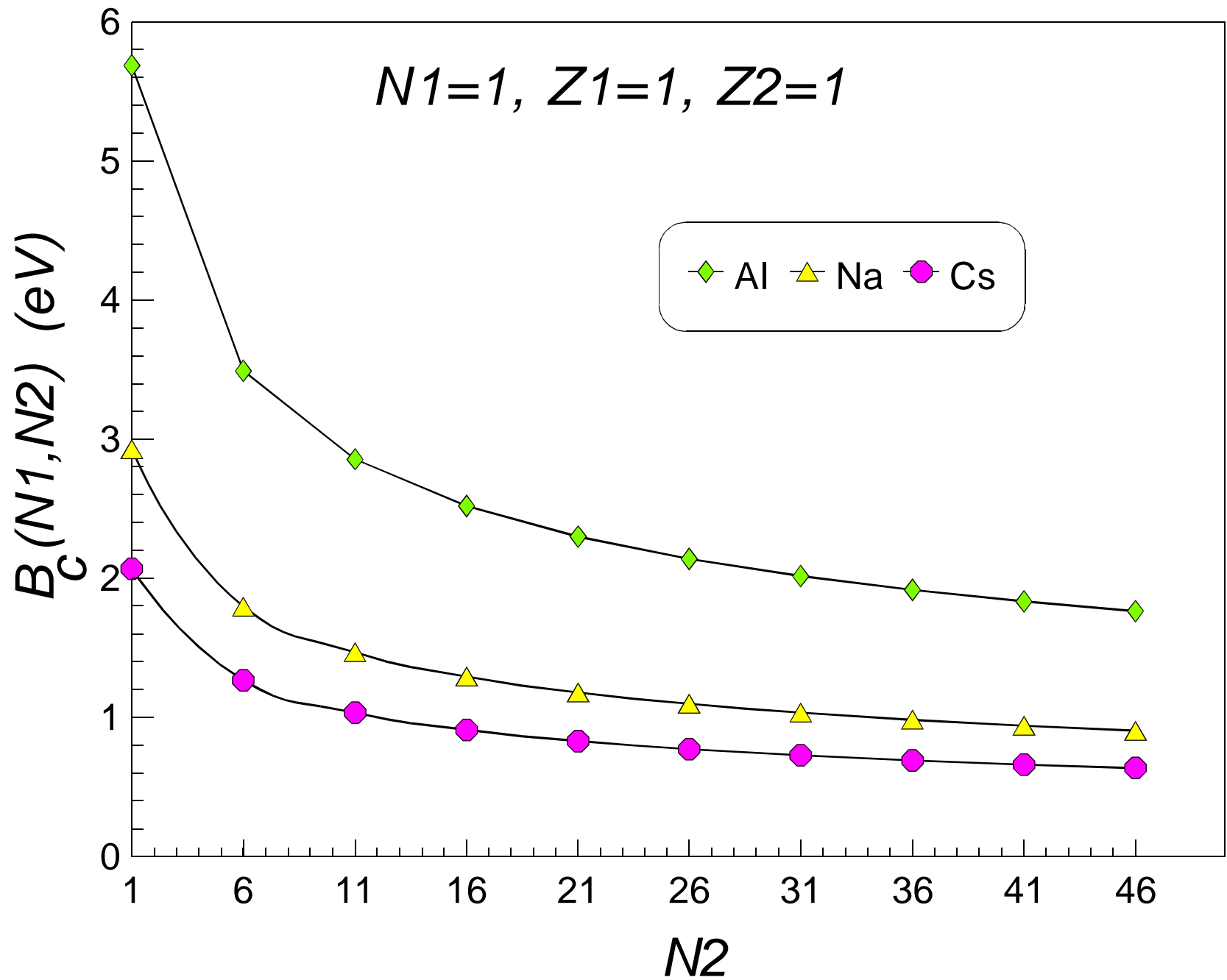
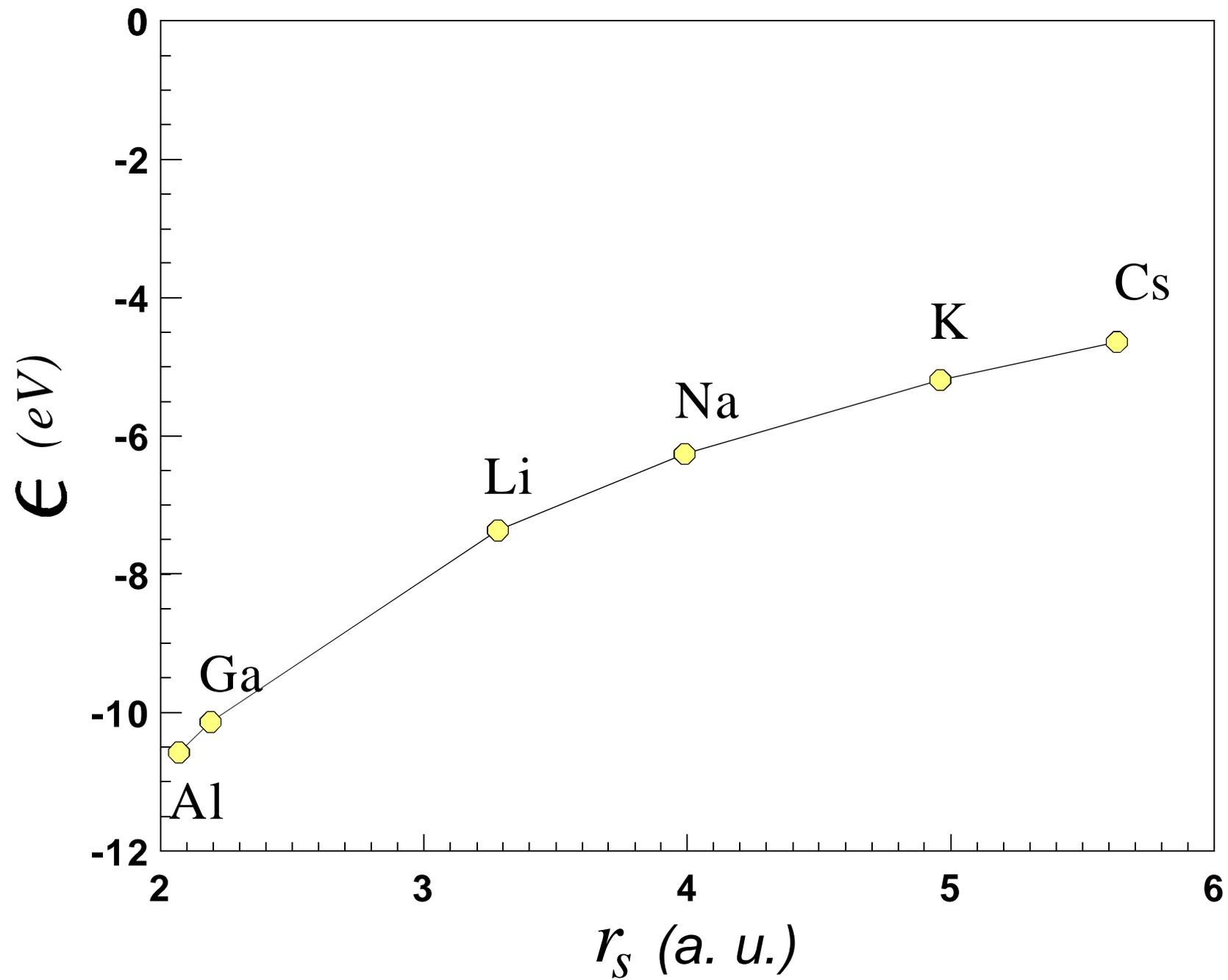
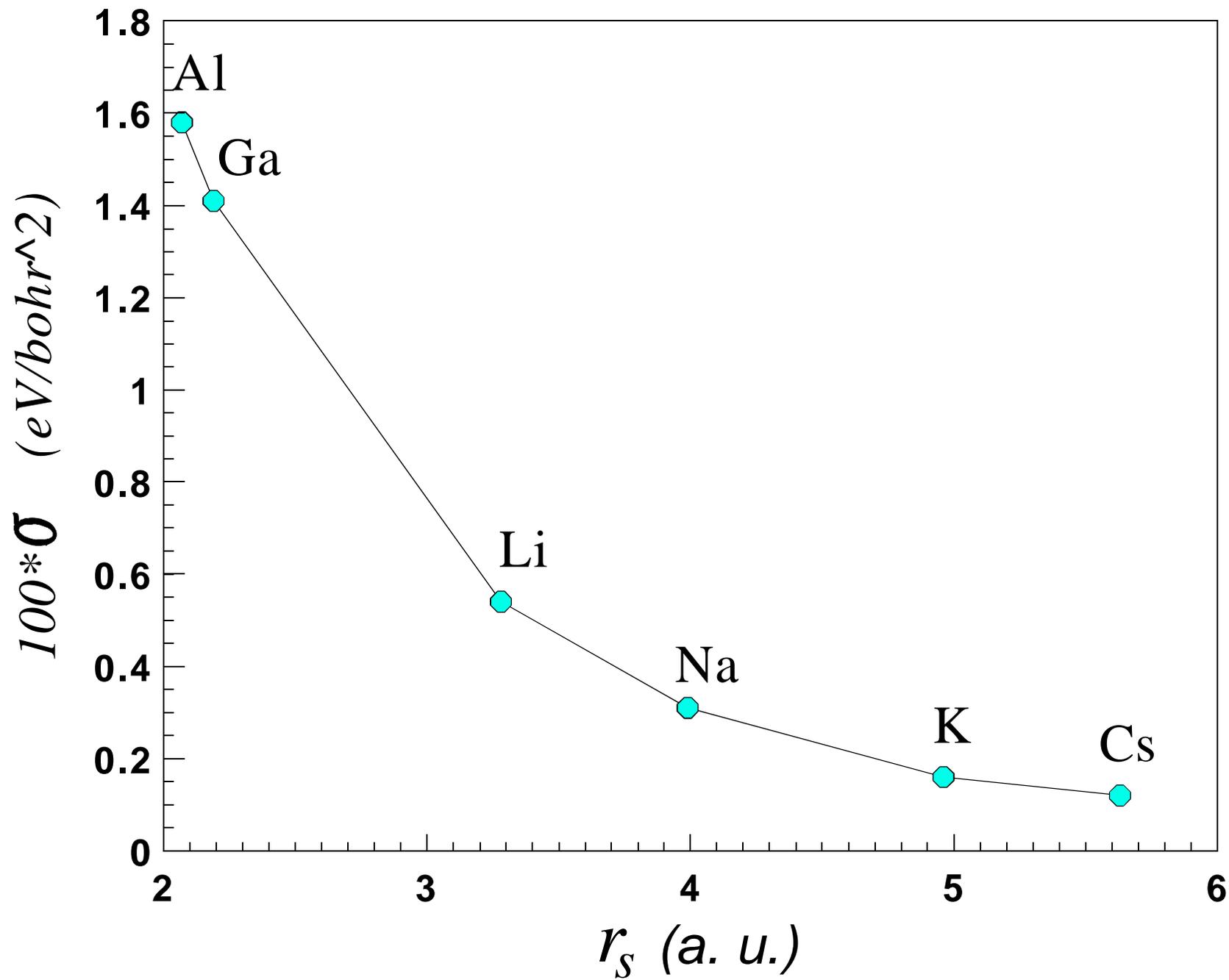
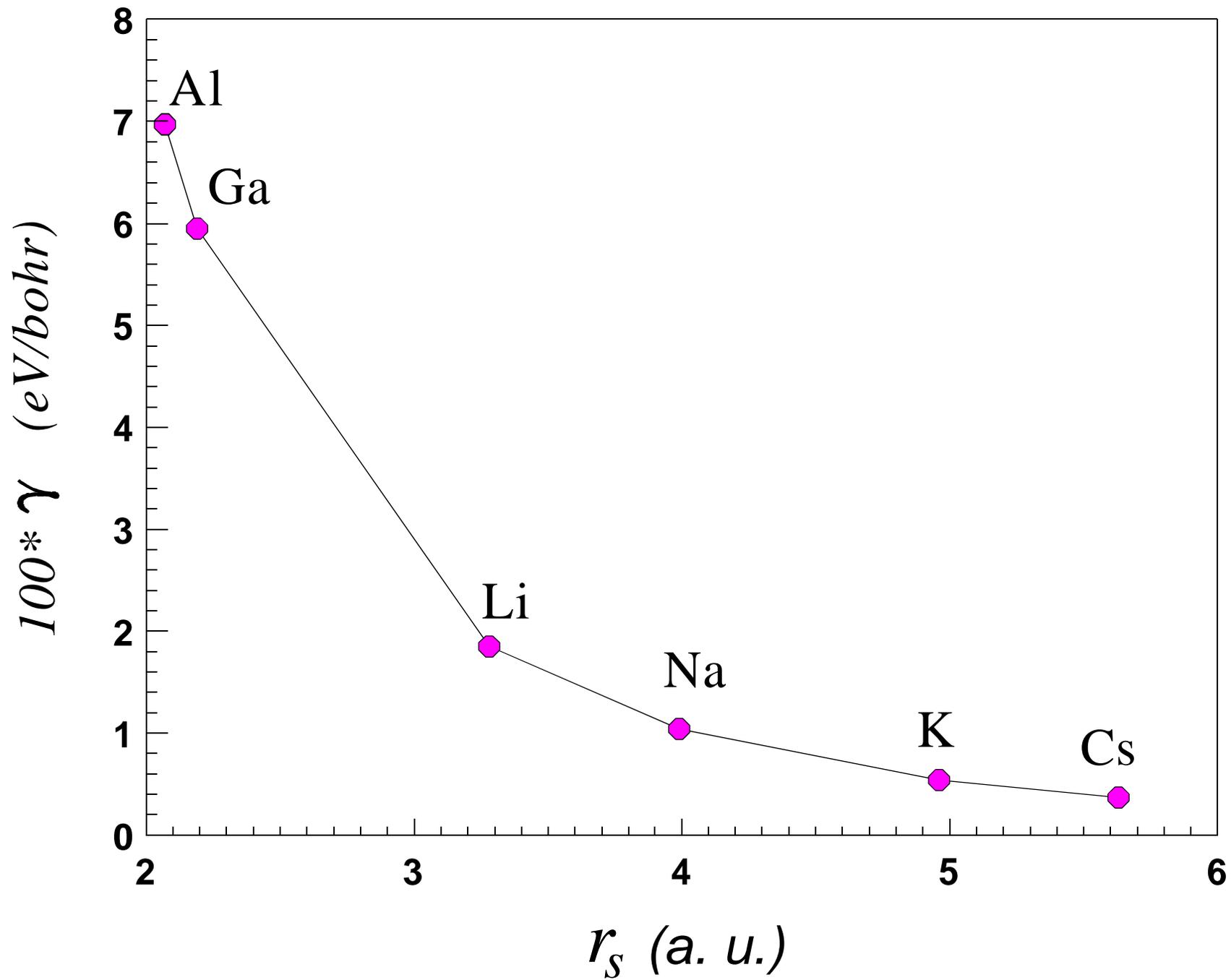
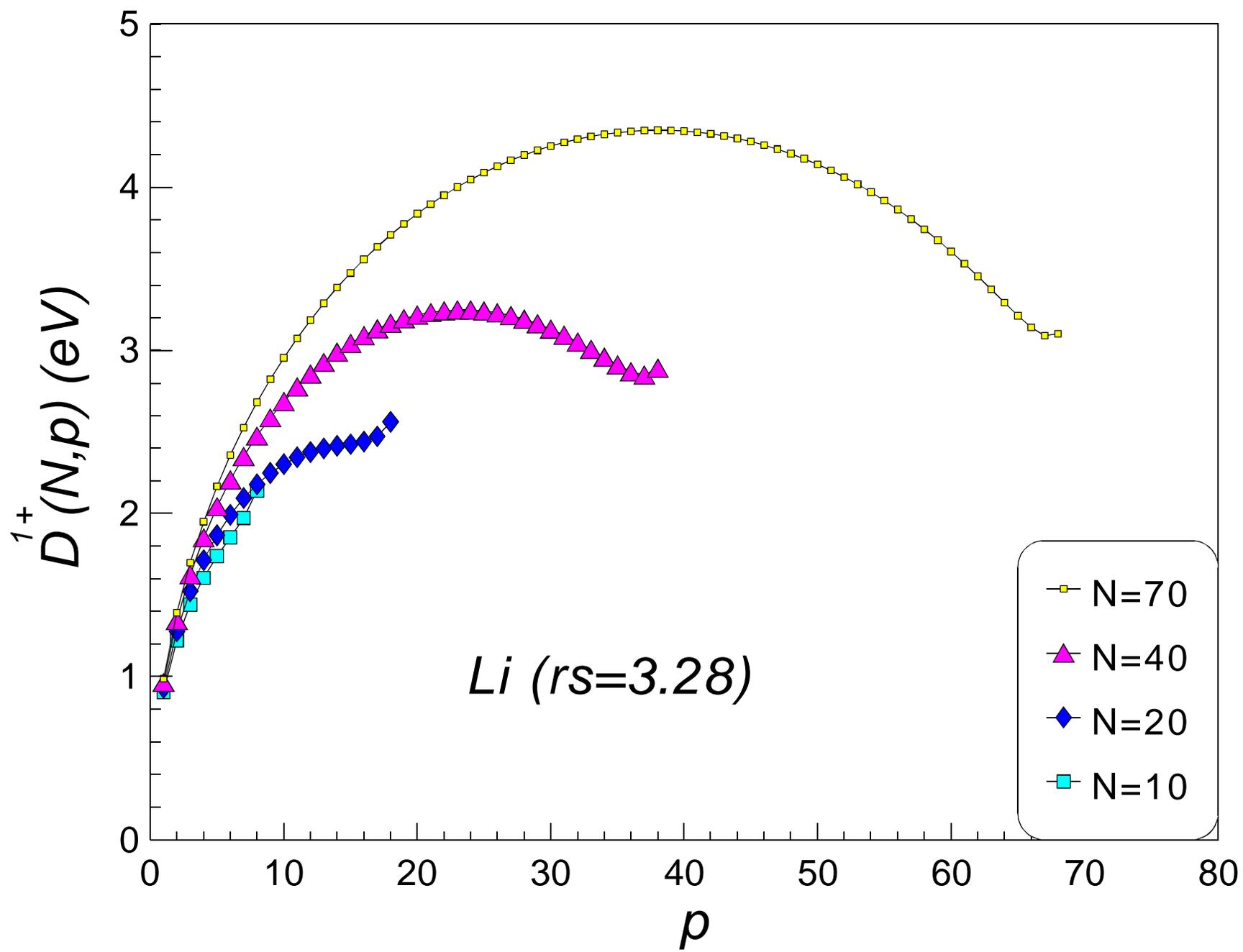
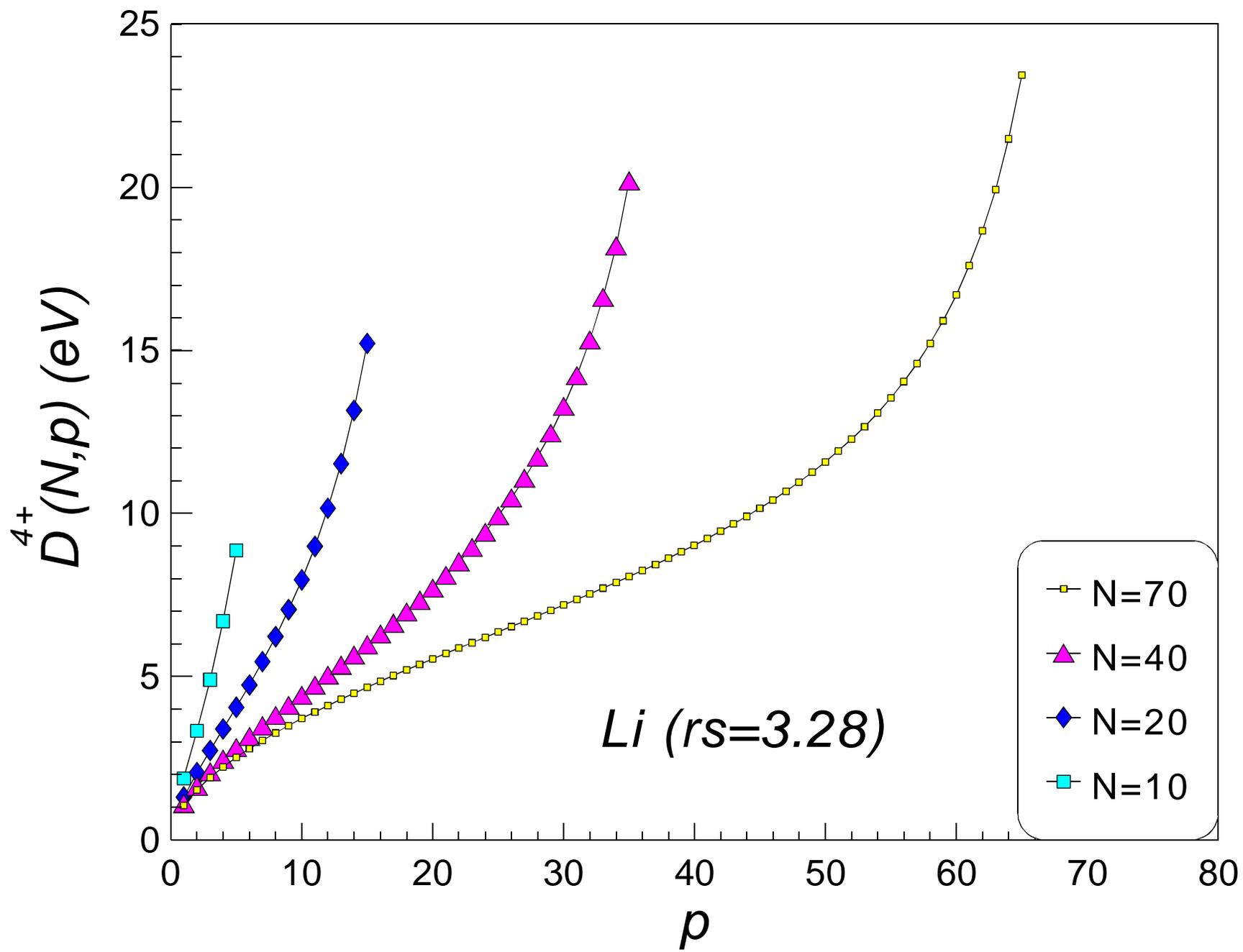
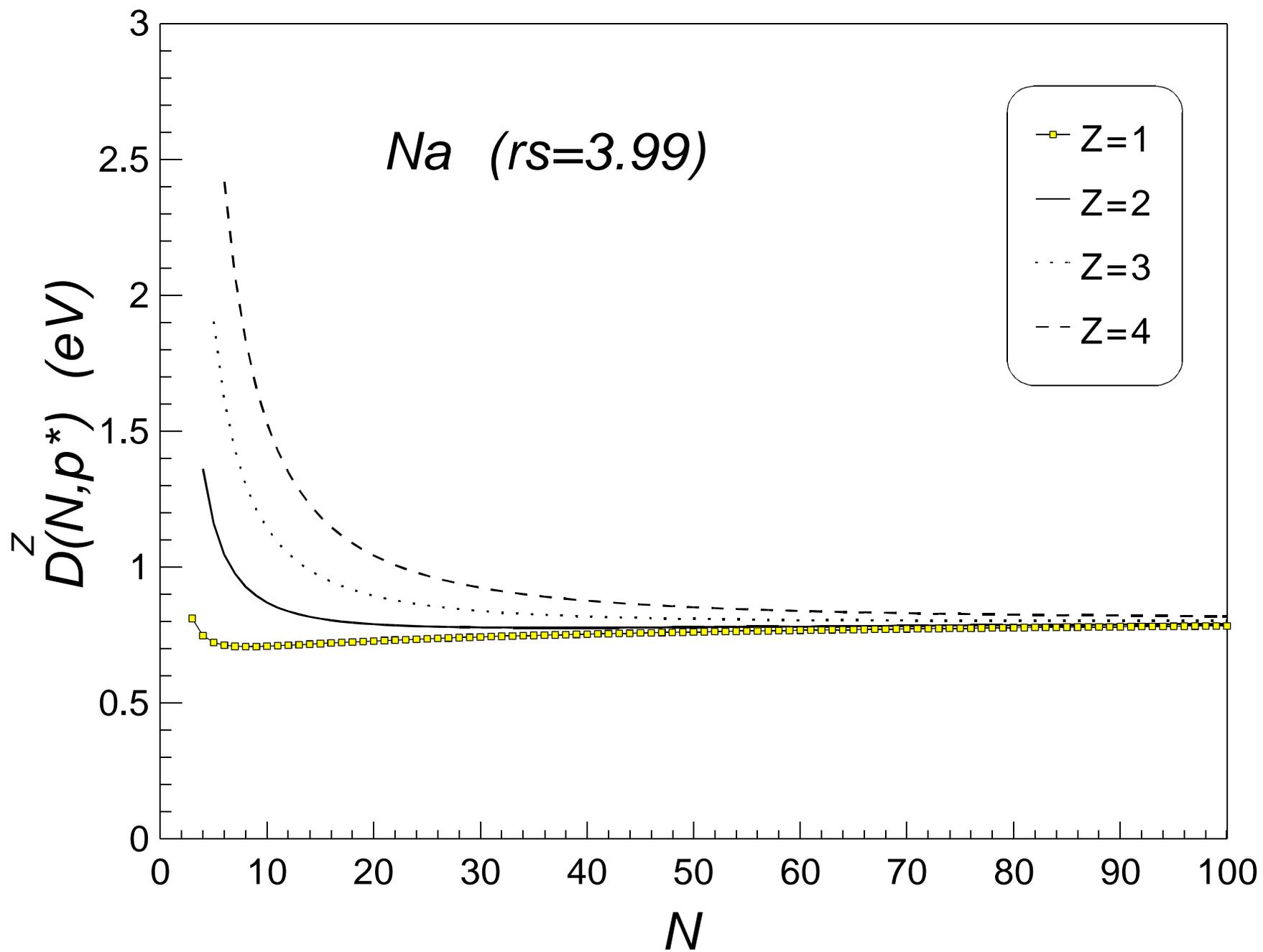
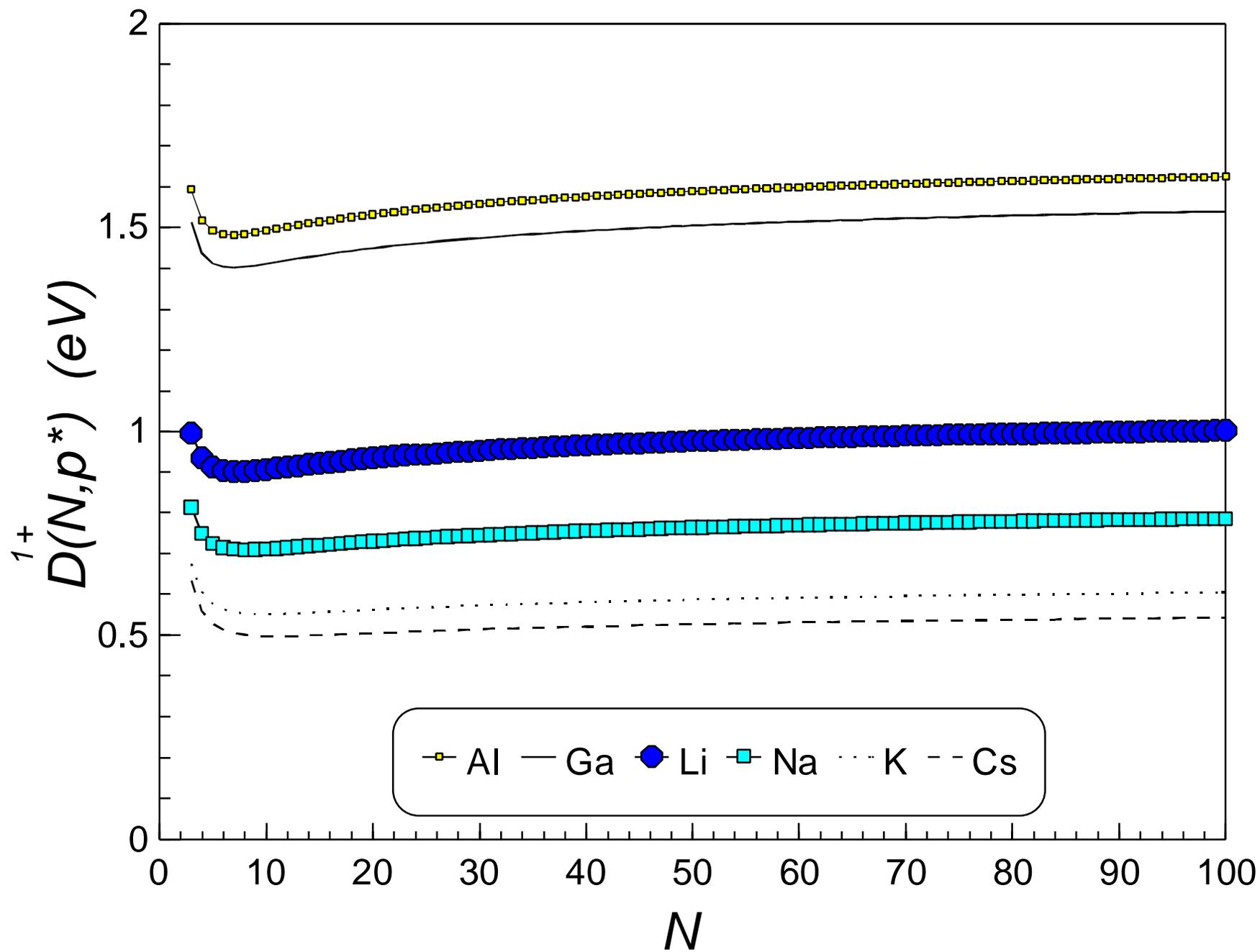
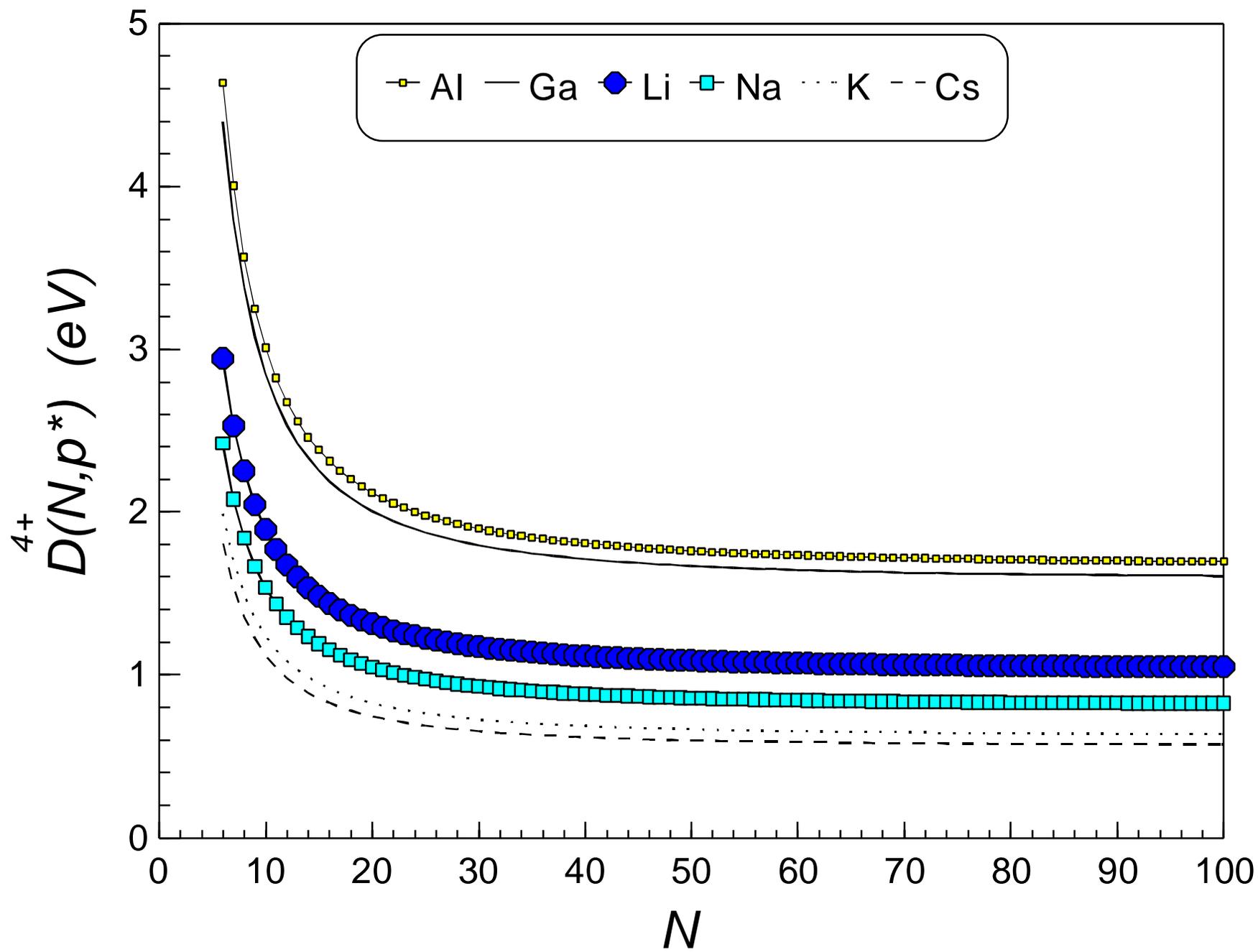
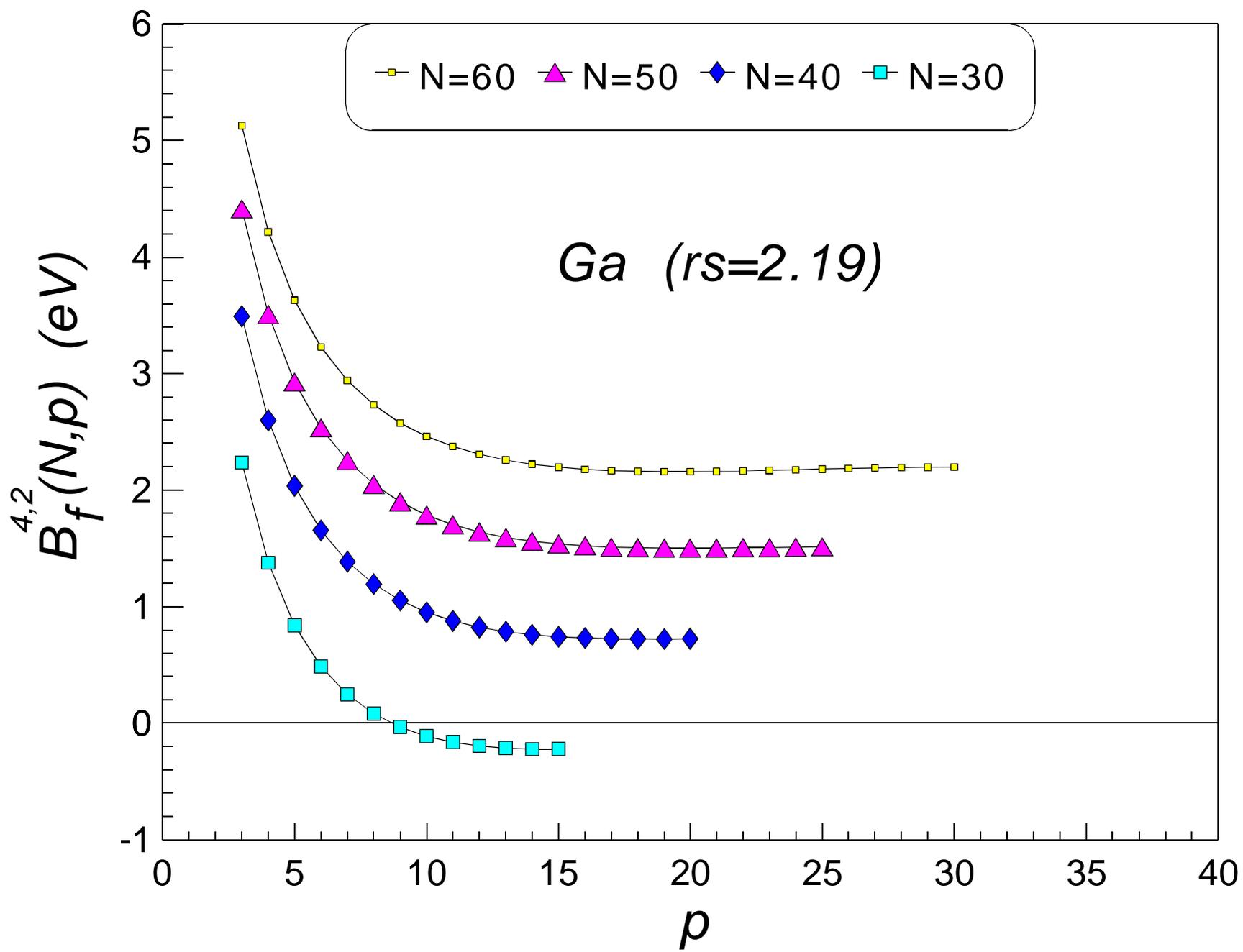
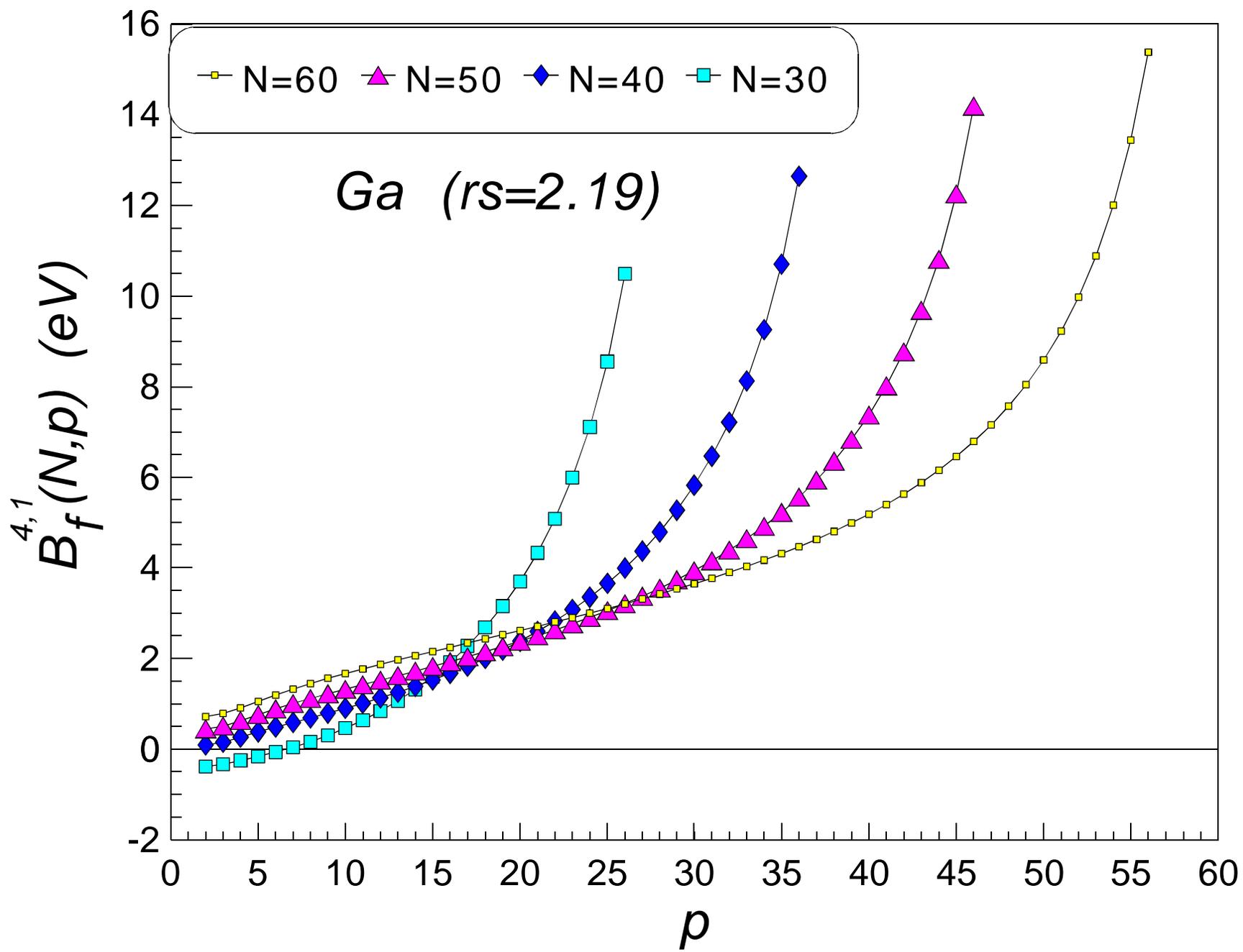
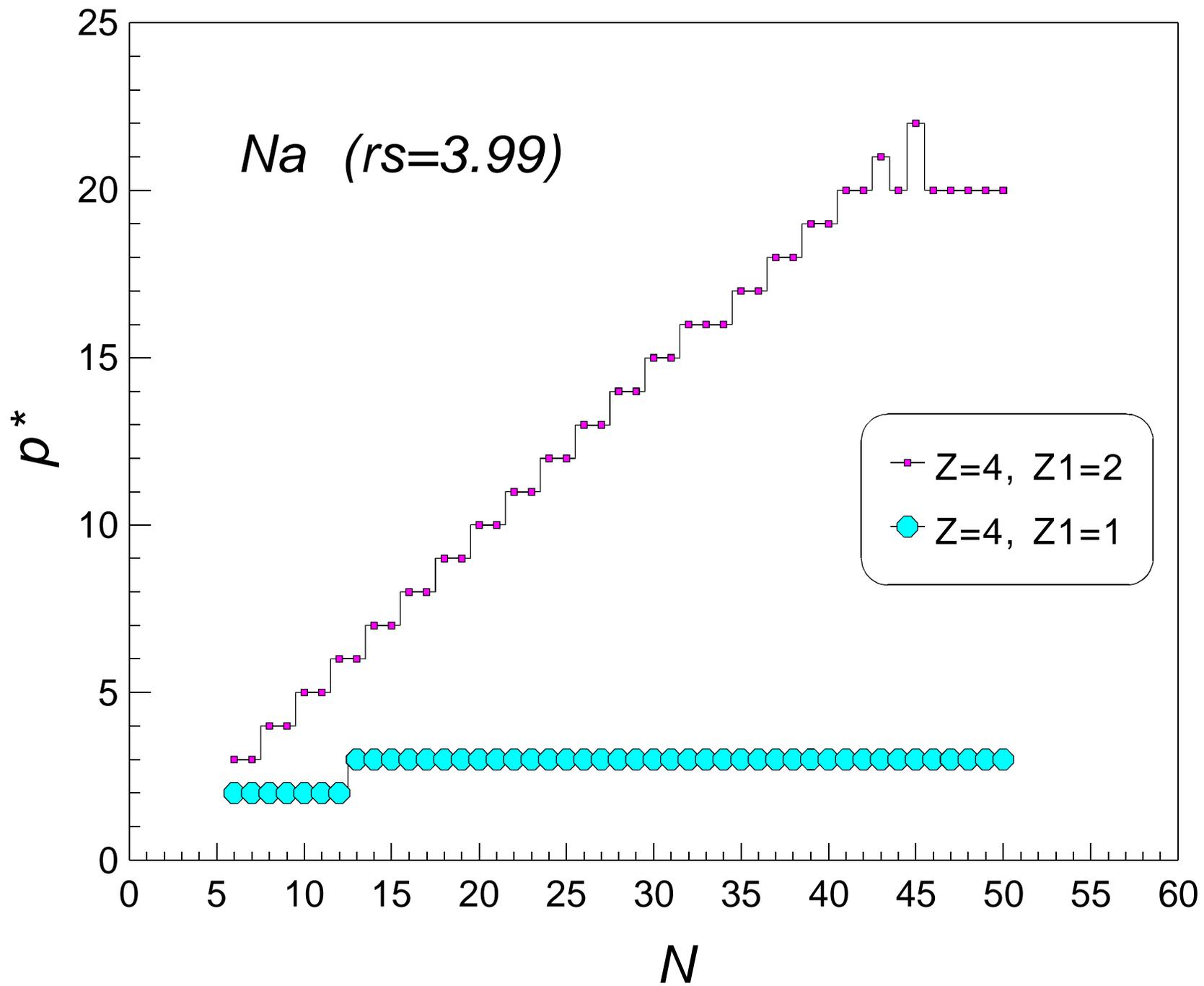
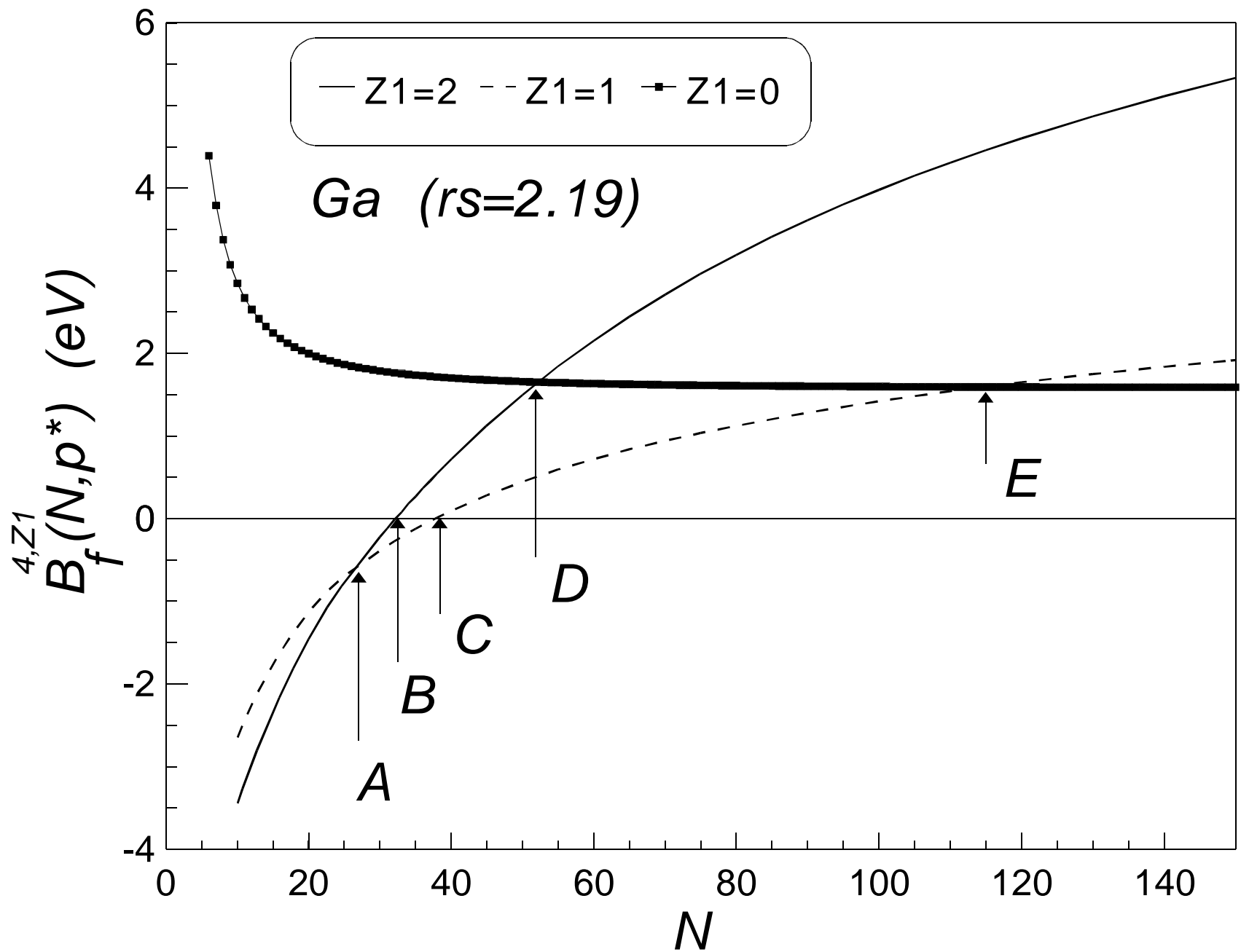
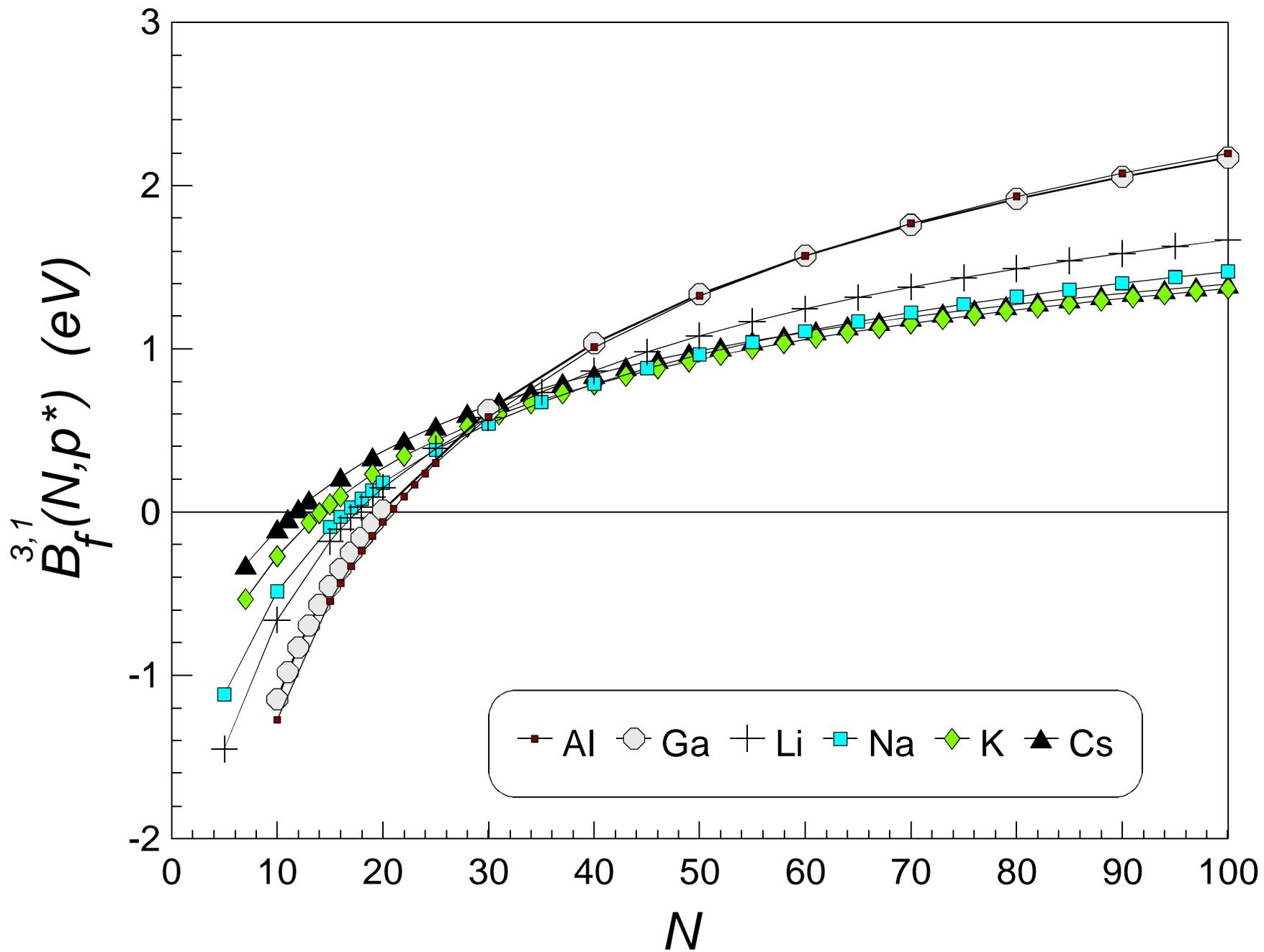
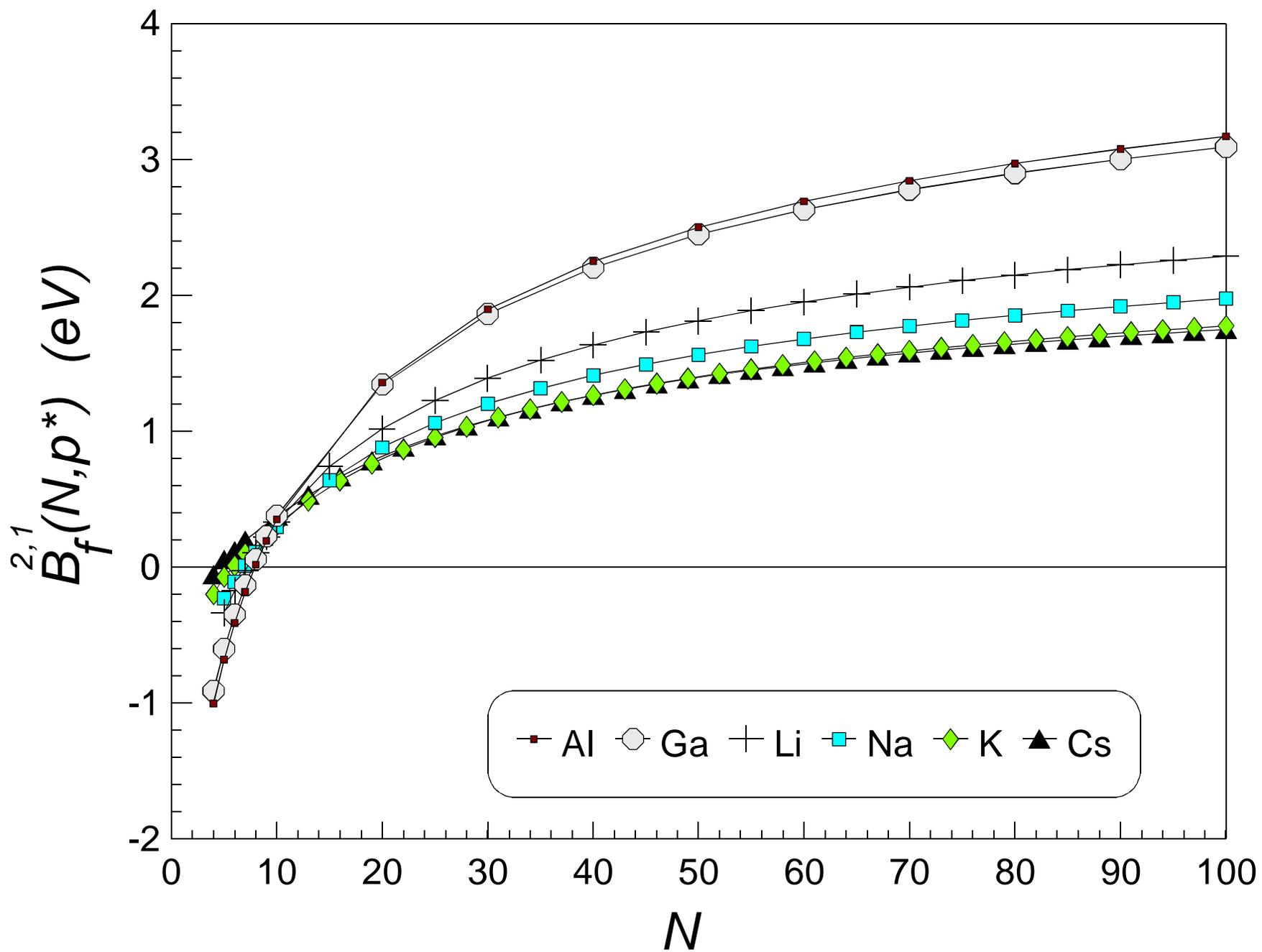
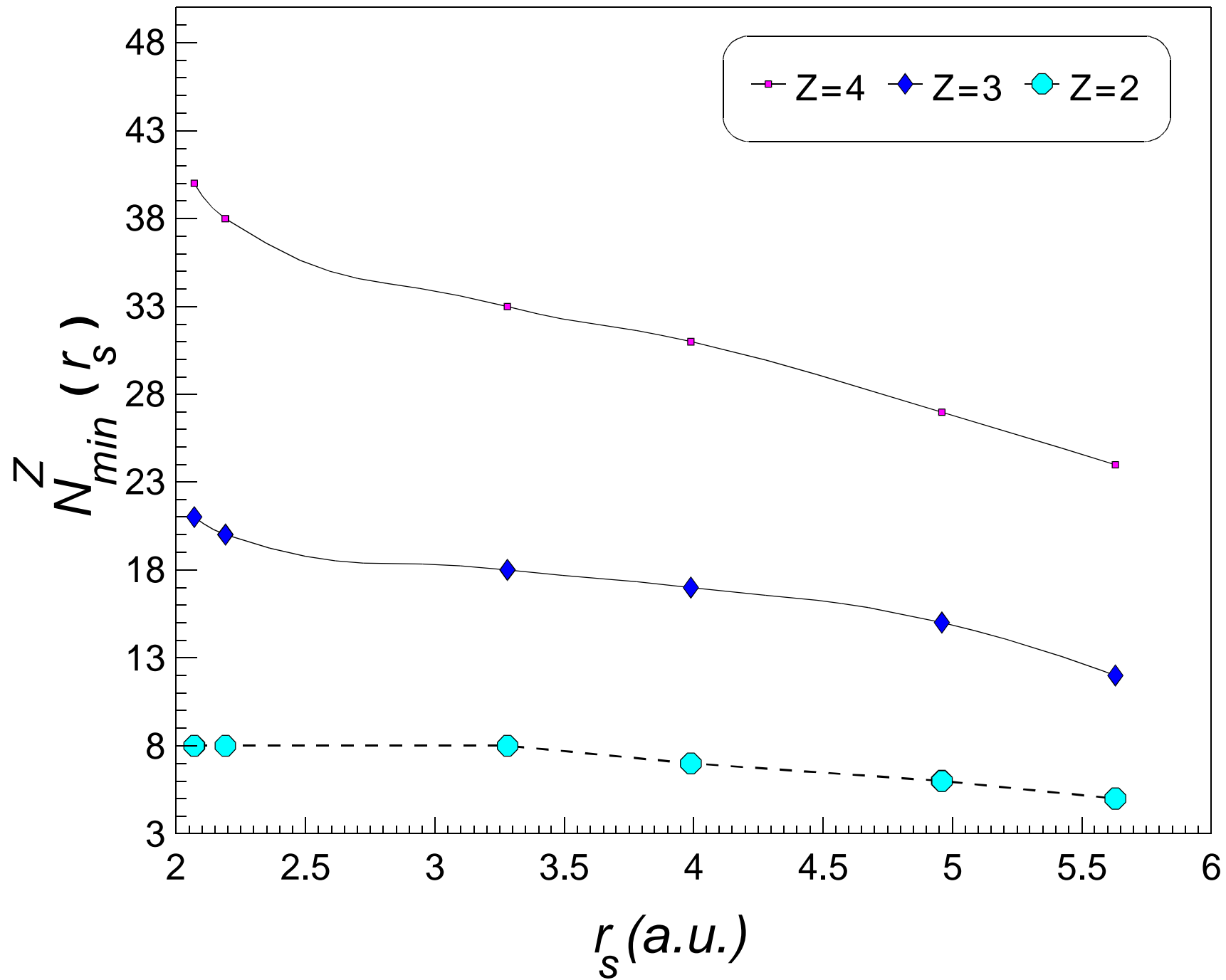
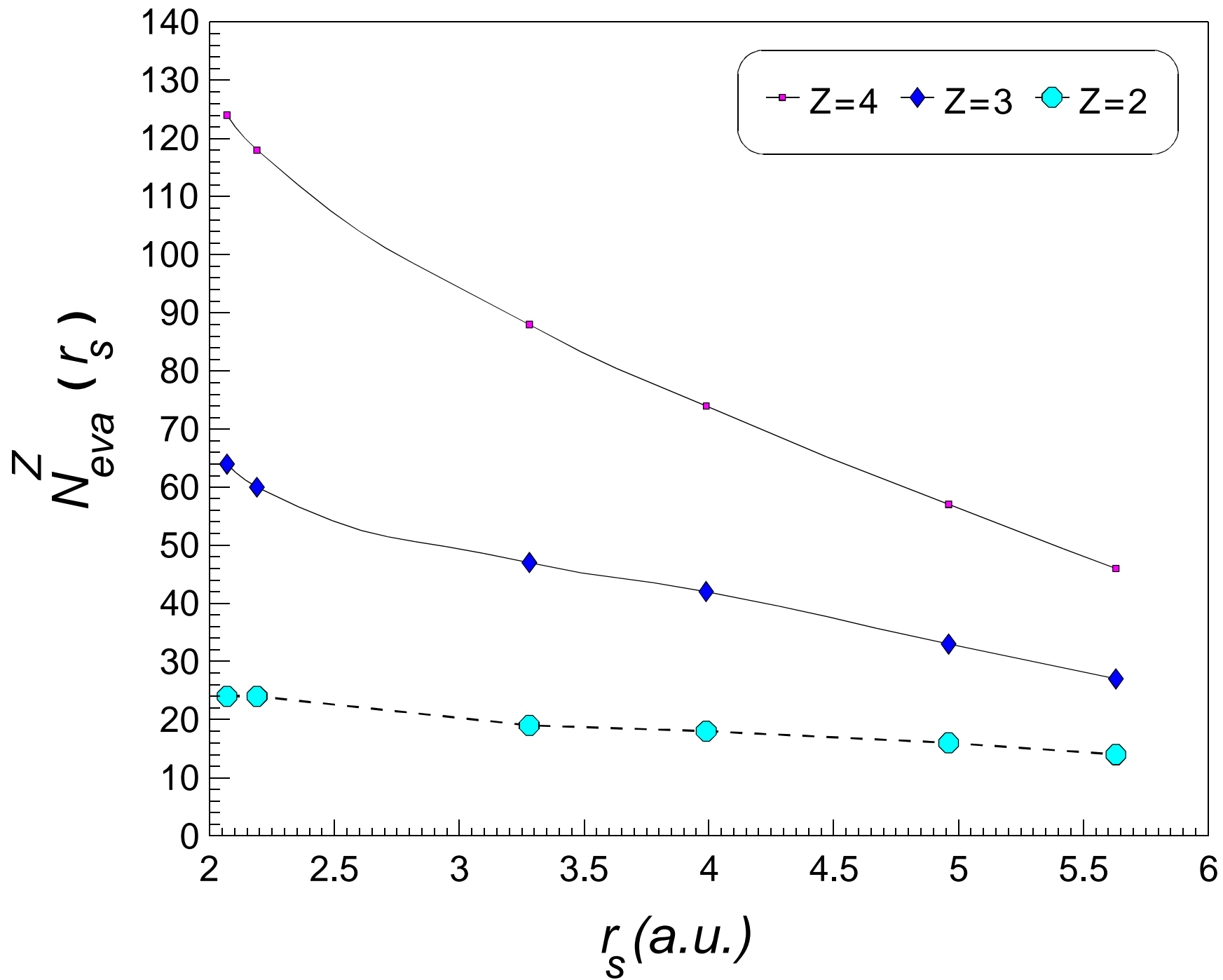
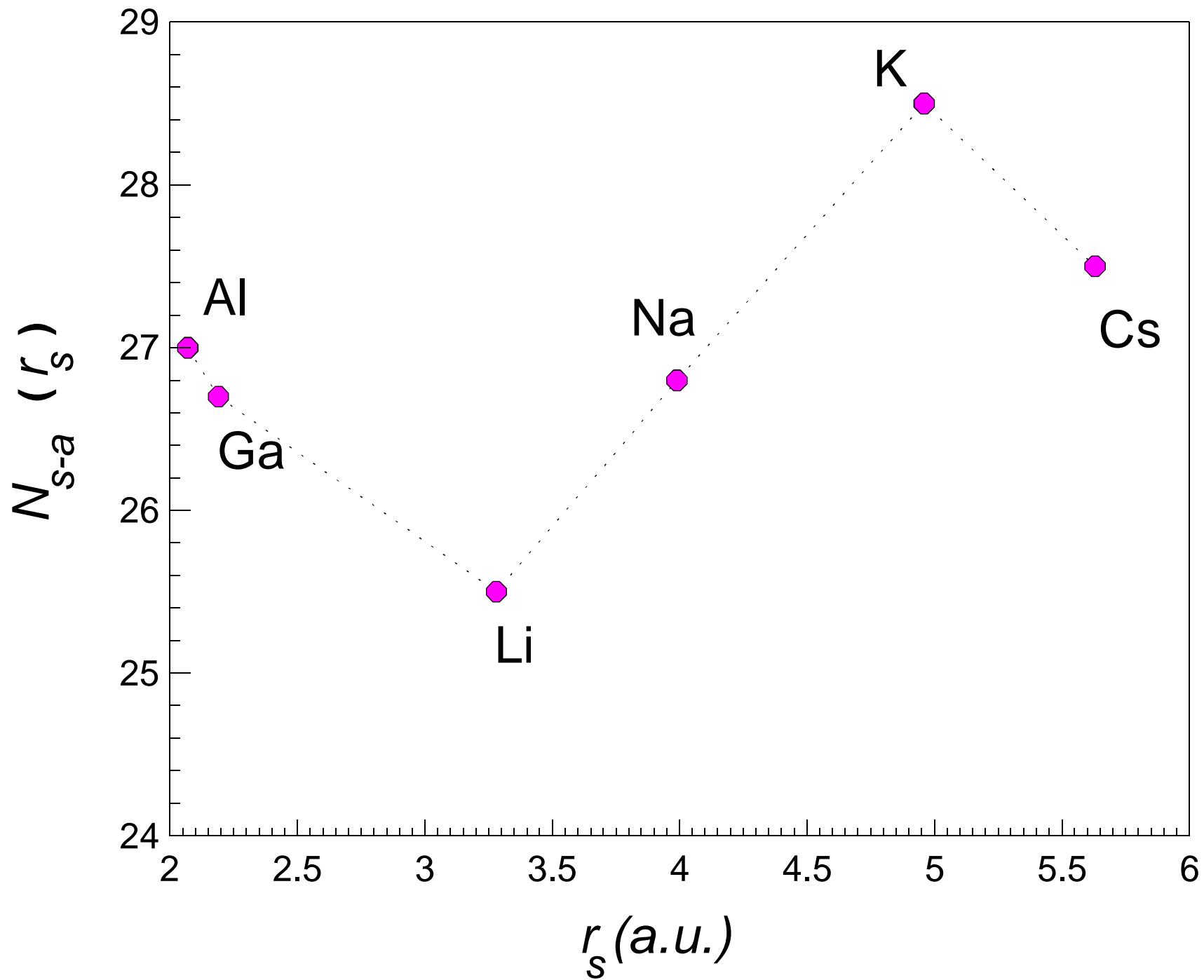
\end{document}